\documentclass[a4paper,fleqn,usenatbib]{mnras}

\usepackage[T1]{fontenc}
\usepackage{ae,aecompl}

\usepackage{graphicx,subfig}	
\usepackage{amsmath}	
\usepackage{amssymb}	
\usepackage{xspace}
\usepackage{scrextend}

\usepackage{natbib,bm}
\usepackage[usenames]{color}
\usepackage{textcomp}
\usepackage[para]{threeparttable}

\newcommand{\bd}{\begin{displaymath}}
\newcommand{\ed}{\end{displaymath}}
\newcommand{\be}{\begin{equation}}
\newcommand{\ee}{\end{equation}}
\newcommand{\beaa}{\begin{eqnarray*}}
\newcommand{\eeaa}{\end{eqnarray*}}
\newcommand{\bea}{\begin{eqnarray}}
\newcommand{\eea}{\end{eqnarray}}

\def\hequad{HE\,0435$-$1223}

\def\wfilens{WFI2033$-$4723}
\def\blens{B1608$+$656}
\def\rxjlens{RXJ1131$-$1231}
\def\sdsslens{SDSS 1206$+$4332}
\def\pglens{PG 1115$+$080}

\def\Ok{\Omega_{\rm k}}
\def\Ode{\Omega_{\rm DE}}
\def\Om{\Omega_{\rm m}}
\def\OL{\Omega_{\Lambda}}

\def\tdist{D_{\Delta t}}
\def\tdistmod{D_{\Delta t}^{\rm model}}
\def\Dd{D_{\rm d}}
\def\Dds{D_{\rm ds}}
\def\Ds{D_{\rm s}}

\def\kext{\kappa_{\rm ext}}

\def\zd{z_{\rm d}}
\def\zs{z_{\rm s}}

\def\hst{\textit{HST}}

\def\kmsMpc {\rm km\,s^{-1}\,Mpc^{-1}}

\newcommand{\sref}[1]{Section~\ref{#1}}

\newcommand{\eref}[1]{Equation~(\ref{#1})}

% ...
\usepackage{soul} 
\definecolor{darkgreen}{rgb}{0.0, 0.5, 0.0}

\newcommand{\ulcdm}{73.3_{-1.8}^{+1.7}~\mathrm{km~s^{-1}~Mpc^{-1}}}
\newcommand{\uolcdm}{74.4_{-2.3}^{+2.1}~\mathrm{km~s^{-1}~Mpc^{-1}}}
\newcommand{\uolcdmOk}{0.26_{-0.25}^{+0.17}}
\newcommand{\uwcdm}{81.6_{-5.3}^{+4.9}~\mathrm{km~s^{-1}~Mpc^{-1}}}
\newcommand{\uwcdmw}{-1.90_{-0.41}^{+0.56}}
\newcommand{\uwwacdm}{81.3_{-5.4}^{+5.1}~\mathrm{km~s^{-1}~Mpc^{-1}}}
\newcommand{\hprec}{2.4}
\newcommand{\lenstension}{3.1}
\newcommand{\tension}{5.3}

\newcommand{\pycs}{{\tt PyCS}\xspace}

\title[H0LiCOW XIII: A 2.4\% measurement of $H_{0}$]{H0LiCOW XIII. A $2.4\%$ measurement of $H_{0}$ from lensed quasars: $5.3\sigma$  tension between early and late-Universe probes}

\author[K. C. Wong et al.]{\parbox{\textwidth}{
Kenneth C. Wong,$^{1,2}$\thanks{E-mail: ken.wong@ipmu.jp}
Sherry H. Suyu,$^{3,4,5}$
Geoff C.-F. Chen,$^{6}$
Cristian E. Rusu,$^{2,7,6}$
Martin Millon,$^{8}$
Dominique Sluse,$^{9}$
Vivien Bonvin,$^{8}$
Christopher D. Fassnacht,$^{6}$
Stefan Taubenberger,$^{3}$
Matthew W. Auger,$^{10}$
Simon Birrer,$^{11}$
James H. H. Chan,$^{8}$
Frederic Courbin,$^{8}$
Stefan Hilbert,$^{12,13}$
Olga Tihhonova,$^{8}$
Tommaso Treu,$^{11}$
Adriano Agnello,$^{14}$
Xuheng Ding,$^{11}$
Inh Jee,$^{3}$
Eiichiro Komatsu,$^{3,1}$
Anowar J. Shajib,$^{11}$
Alessandro Sonnenfeld,$^{15}$
Roger D. Blandford,$^{16}$
L\'{e}on V. E. Koopmans,$^{17}$
Philip J. Marshall,$^{16}$
and Georges Meylan$^{8}$
}
\\
\\
\parbox{\textwidth}{
$^{1}$Kavli IPMU (WPI), UTIAS, The University of Tokyo, Kashiwa, Chiba 277-8583, Japan\\
$^{2}$National Astronomical Observatory of Japan, 2-21-1 Osawa, Mitaka, Tokyo 181-8588, Japan\\
$^{3}$Max-Planck-Institut f{\"u}r Astrophysik, Karl-Schwarzschild-Str.~1, 85748 Garching, Germany\\
$^{4}$Physik-Department, Technische Universit\"at M\"unchen, James-Franck-Stra\ss{}e~1, 85748 Garching, Germany\\
$^{5}$Academia Sinica Institute of Astronomy and Astrophysics (ASIAA), 11F of ASMAB, No.1, Section 4, Roosevelt Road, Taipei 10617, Taiwan\\
$^{6}$Department of Physics, University of California, Davis, CA 95616, USA\\
$^{7}$Subaru Telescope, National Astronomical Observatory of Japan, 650 N Aohoku Pl, Hilo, HI 96720, USA\\
$^{8}$Institute of Physics, Laboratory of Astrophysics, Ecole Polytechnique F\'{e}d\'{e}rale de Lausanne (EPFL), Observatoire de Sauverny, 1290 Versoix, Switzerland\\
$^{9}$STAR Institute, Quartier Agora - All\'{e}e du six Ao\^{u}t, 19c B-4000 Li\`{e}ge, Belgium\\
$^{10}$Institute of Astronomy, University of Cambridge, Madingley Road, Cambridge CB3 0HA, UK\\
$^{11}$Department of Physics and Astronomy, University of California, Los Angeles, CA 90095, USA\\
$^{12}$Exzellenzcluster Universe, Boltzmannstr. 2, D-85748 Garching, Germany\\
$^{13}$Ludwig-Maximilians-Universit\"at, Universit\"ats-Sternwarte, Scheinerstr. 1, D-81679 M\"unchen, Germany\\
$^{14}$DARK, Niels-Bohr Institute, Lyngbyvej 2, 2100 Copenhagen, Denmark\\
$^{15}$Leiden Observatory, Leiden University, Niels Bohrweg 2, 2333 CA Leiden, the Netherlands\\
$^{16}$Kavli Institute for Particle Astrophysics and Cosmology, Stanford University, 452 Lomita Mall, Stanford, CA 94035, USA\\
$^{17}$Kapteyn Astronomical Institute, University of Groningen, PO Box 800, NL-9700 AV Groningen, The Netherlands\\
}}

\date{Accepted XXX. Received YYY; in original form ZZZ}

\pubyear{2019}

\begin{document}
\label{firstpage}
\pagerange{\pageref{firstpage}--\pageref{lastpage}}
\maketitle


\begin{abstract}
We present a measurement of the Hubble constant ($H_{0}$) and other cosmological parameters from a joint analysis of six gravitationally lensed quasars with measured time delays.  All lenses except the first are analyzed blindly with respect to the cosmological parameters.  In a flat $\Lambda$CDM cosmology, we find $H_{0} = \ulcdm$, a $\hprec\%$ precision measurement, in agreement with local measurements of $H_{0}$ from type Ia supernovae calibrated by the distance ladder, but in $\lenstension\sigma$ tension with {\it Planck} observations of the cosmic microwave background (CMB).  This method is completely independent of both the supernovae and CMB analyses.  A combination of time-delay cosmography and the distance ladder results is in $\tension\sigma$ tension with {\it Planck} CMB determinations of $H_{0}$ in flat $\Lambda$CDM.  We compute Bayes factors to verify that all lenses give statistically consistent results, showing that we are not underestimating our uncertainties and are able to control our systematics.  We explore extensions to flat $\Lambda$CDM using constraints from time-delay cosmography alone, as well as combinations with other cosmological probes, including CMB observations from {\it Planck}, baryon acoustic oscillations, and type Ia supernovae.  Time-delay cosmography improves the precision of the other probes, demonstrating the strong complementarity.  Allowing for spatial curvature does not resolve the tension with {\it Planck}.  Using the distance constraints from time-delay cosmography to anchor the type Ia supernova distance scale, we reduce the sensitivity of our $H_0$ inference to cosmological model assumptions.  For six different cosmological models, our combined inference on $H_0$ ranges from $\sim73$--$78~\mathrm{km~s^{-1}~Mpc^{-1}}$, which is consistent with the local distance ladder constraints.

\end{abstract}

\begin{keywords}
cosmology: observations $-$
cosmology: cosmological parameters $-$
distance scale $-$
gravitational lensing: strong
\end{keywords}


\section{Introduction} \label{sec:intro}
The flat $\Lambda$ cold dark matter ($\Lambda$CDM) cosmological model has proven to be remarkably successful at describing the Universe as measured by a wide range of experiments, particularly observations of the cosmic microwave background (CMB).  The final results from the {\it Planck} mission \citep{planck+2018a} provide the most precise constraints on cosmological parameters to date from CMB observations \citep{planck+2018b}.  However, relaxing the flat $\Lambda$CDM assumption by introducing additional complexities, such as non-zero curvature, an equation of state parameter $w \ne -1$, or a time-varying $w$, leads to much weaker constraints and large degeneracies among the various cosmological parameters.  In particular, many parameters become degenerate with the Hubble constant, $H_{0}$, which sets the present-day expansion rate of the universe.  $H_{0}$ cannot be constrained directly from CMB observations, but must be inferred by first assuming a cosmological model.  In this context, measuring $H_{0}$ independent of CMB observations is one of the most important complementary probes for understanding the nature of the Universe \citep{weinberg+2013}.

The most well-established method for measuring $H_{0}$ is through observations of type Ia supernovae (SNe).  Type Ia SNe are ``standardizable candles" in that their luminosities, and thus their absolute distances, can be determined by the evolution of their light curves, and therefore can be used to infer $H_{0}$ from the slope of their distance-redshift relation.  Type Ia SNe luminosities are typically calibrated via the ``distance ladder" \citep[e.g.,][]{sandage+2006,freedman+2012,riess+2016,riess+2018,riess+2019}, in which parallax measurements are used to determine distances to nearby Cepheid variable stars (which have a known period-luminosity relation) and are in turn used to determine distances to type Ia SNe in the Hubble flow.

Recent determinations of $H_{0}$ from the Supernovae, $H_{0}$, for the Equation of State of Dark Energy \citep[SH0ES;][]{riess+2016} collaboration using this method are in tension with the {\it Planck} CMB measurements under the flat $\Lambda$CDM model \citep[e.g.,][]{bernal+2016,freedman2017}.  The latest SH0ES result finds $H_{0} = 74.03 \pm 1.42~\mathrm{km~s^{-1}~Mpc^{-1}}$ \citep{riess+2019}, which differs from the {\it Planck} flat $\Lambda$CDM result\footnote{Baseline $\Lambda$CDM chains with baseline likelihoods (based on plikHM\_TTTEEE\_lowl\_lowE)} of $H_{0} = 67.4 \pm 0.5~\mathrm{km~s^{-1}~Mpc^{-1}}$ \citep{planck+2018b} by $4.4\sigma$.  Possible systematic errors in one or both methods may resolve this tension \citep[e.g.,][]{rigault+2015,rigault+2018}, but investigations thus far have yet to conclusively identify any such systematic \citep[e.g.,][]{addison+2018,jones+2018,roman+2018,camarenamarra2019,rose+2019}.  Furthermore, independent determinations of $H_{0}$ using the ``inverse distance ladder" method \citep[e.g.,][]{aubourg+2015,cuesta+2015,macaulay+2019} are in agreement with the {\it Planck} value, although this depends on assumptions of the physical scale of the sound horizon \citep[e.g.,][]{aylor+2019,macaulay+2019,arendse+2019a,arendse+2019b}.  Other methods, such as CMB polarization measurements \citep[e.g.,][]{henning+2018}, galaxy clustering \citep[e.g.,][]{abbott+2018a}, water masers \citep[e.g.,][]{herrnstein+1999,humphreys+2013,braatz+2018}, X-ray observations of SZ galaxy clusters \citep[e.g.,][]{silkwhite1978,reese+2002,bonamente+2006,kozmanyan+2019}, the Balmer line $L-\sigma$ relation of HII galaxies \citep[e.g.,][]{melnick+2000,chavez+2012,gonzalezmoran+2019}, extragalactic background light attenuation \citep[e.g.,][]{salamon+1994,dominguezprada2013,dominguez+2019}, type IIP supernova expanding photospheres \citep[e.g.,][]{schmidt+1994, gall+2016}, and gravitational waves \citep[e.g.,][]{abbott+2017,feeney+2019,soaressantos+2019}, have yet to resolve the $H_{0}$ discrepancy, as their precision is not yet comparable to {\it Planck} or SH0ES, or they require additional assumptions.  If unresolved, this tension may force the rejection of the flat $\Lambda$CDM model and indicate new physics that must be incorporated into our understanding of cosmology.

After the submission of this paper, an alternate calibration of the distance ladder using the ``tip of the red giant branch" (TRGB) method by the Carnegie-Chicago Hubble Program \citep[CCHP;][]{beaton+2016} found an intermediate value of $H_{0} = 69.8 \pm 1.9~\mathrm{km~s^{-1}~Mpc^{-1}}$ \citep{freedman+2019}.  However, this measurement is not fully independent of SH0ES since they share some calibrating sources (i.e., galaxies hosting SNe that are close enough for Cepheid and/or TRGB distance measurements), and there is an ongoing debate about the results from this method \citep[e.g.,][]{yuan+2019}, further highlighting the need for additional independent probes (see \citealp{verde+2019} for a recent review of the field).

Gravitational lensing offers an independent method of determining $H_{0}$.  When a background object (the ``source") is gravitationally lensed into multiple images by an intervening mass (the ``lens"), light rays emitted from the source will take different paths through space-time at the different image positions.  Because these paths have different lengths and pass through different gravitational potentials, light rays emitted from the source at the same time will arrive at the observer at different times depending on which image it arrives at.  If the source is variable, this ``time delay" between multiple images can be measured by monitoring the lens and looking for flux variations corresponding to the same source event.  The time delay is related to a quantity referred to as the ``time-delay distance", $\tdist$, and depends on the mass distribution in the lensing object, the mass distribution along the line of sight (LOS), and cosmological parameters.  $\tdist$ is primarily sensitive to $H_{0}$, although there is a weak dependence on other parameters \citep[e.g.,][]{coemoustakas2009,linder2011,treumarshall2016}.  This one-step method is completely independent of and complementary to the CMB and the distance ladder.  The distances probed by time-delay cosmography are also larger than those from the distance ladder, making this method immune to a monopole in the bulk velocity field of the local Universe (i.e., a ``Hubble bubble").

This method of using gravitational lens time delays to measure $H_{0}$ was first proposed by \citet{refsdal1964}, who suggested using lensed SNe for this purpose.  In practice, finding lensed SNe with resolved images is extremely rare, with only two such lenses having been discovered to date \citep{kelly+2015,goobar+2017}.  While the prospect of discovering more lensed SNe in future imaging surveys and measuring their time delays is promising \citep[e.g.,][]{ogurimarshall2010,goldsteinnugent2017,goldstein+2018,huber+2019,wojtak+2019}, lensed quasars have generally been used to constrain $H_{0}$ in this manner \citep[e.g.,][]{vanderriest+1989,keetonkochanek1997,schechter+1997,kochanek2003,koopmans+2003,saha+2006,oguri2007,vuissoz+2008,fadely+2010,suyu+2010,suyu+2013,serenoparaficz2014,rathnakumar+2015,birrer+2016,birrer+2019,chen+2016,wong+2017,bonvin+2017} due to their brightness and variable nature.

Measuring $H_{0}$ from lensed quasars through this method requires a variety of observational data.  Long-term dedicated photometric monitoring of the lens is needed to obtain accurate time delays \citep[e.g.,][]{bonvin+2017,bonvin+2018}.  Several years of monitoring are generally required to overcome microlensing variability, although \citet{courbin+2018} recently demonstrated that delays could be measured from just one year of monitoring owing to high photometric accuracy (milli-mag) and observing cadence (daily).  In addition, deep high-resolution imaging of the lens is required to observe the extended images of the quasar host galaxy, which is needed to break degeneracies in the lens modeling between the mass profile and the underlying cosmology \citep[e.g.,][]{kochanek2002,koopmans+2003,dye+2005}.  Furthermore, to mitigate the effects of the mass-sheet degeneracy \citep[e.g.,][]{falco+1985,gorenstein+1988,saha2000,schneidersluse2013,xu+2016}, it is important to obtain a measurement of the lens galaxy's velocity dispersion \citep[e.g.][]{treukoopmans2002,koopmans+2003,koopmans2004,sonnenfeld2018}.  Finally, observational data to constrain the mass along the LOS to the lens are needed to estimate the external convergence, $\kext$, which can bias the inferred $\tdist$ if unaccounted for \citep[e.g.,][]{collett+2013,greene+2013,mccully+2014,mccully+2017,sluse+2017,rusu+2017,tihhonova+2018}.

The $H_{0}$ Lenses in COSMOGRAIL's Wellspring (H0LiCOW) collaboration  \citep[][hereafter H0LiCOW I]{suyu+2017} has provided the strongest constraints on $H_{0}$ to date from time-delay cosmography.  Our most recent measurements had constrained $H_{0}$ to $3.0\%$ precision for a flat $\Lambda$CDM cosmology from a combination of four lensed quasars \citep[][hereafter H0LiCOW IX]{birrer+2019}.  We attain this precision by taking advantage of our substantial dataset, which includes accurate time-delay measurements from the COSmological MOnitoring of GRAvItational Lenses \citep[COSMOGRAIL;][]{courbin+2005,eigenbrod+2005,bonvin+2018} project and radio-wavelength monitoring \citep{fassnacht+2002}, deep {\it Hubble Space Telescope} ({\it HST}) and/or ground-based adaptive optics (AO) imaging \citep[H0LiCOW I, IX,][]{chen+2016,chen+2019}, spectroscopy of the lens galaxy to measure its velocity dispersion \citep[e.g.,][hereafter H0LiCOW X]{sluse+2019}, and deep wide-field spectroscopy and imaging to characterize the LOS in these systems \citep[e.g.,][hereafter H0LiCOW II and H0LiCOW III, respectively]{sluse+2017, rusu+2017}.

In this milestone paper, we present the latest constraints on $H_{0}$ from H0LiCOW from a combined sample of six lensed quasars.  Two of the four lenses analyzed previously, \hequad~\citep[][hereafter H0LiCOW IV]{wong+2017} and \rxjlens~\citep{suyu+2014}, are reanalyzed using new AO data \citep{chen+2019}.  We add constraints from two newly-analyzed systems -- \pglens~\citep{chen+2019} and \wfilens~\citep[][hereafter H0LiCOW XII]{rusu+2019} -- to provide the tightest $H_{0}$ constraints to date from time-delay cosmography.

This paper is organized as follows.  In Section~\ref{sec:tdcosmo}, we summarize the theory behind using time-delay cosmography to infer the time-delay distance, which is inversely proportional to $H_{0}$.  In Section~\ref{sec:analysis}, we present our lens sample and describe how our data and analysis methods allow us to constrain $H_{0}$ to a precision greater than what has previously been possible from time-delay strong lensing.  In Section~\ref{sec:combine_lenses}, we verify that our lenses are consistent with each other so that we can combine them for our cosmological inference.  We present our main results for flat $\Lambda$CDM and more complex cosmologies in Section~\ref{sec:cosmo}.  In Section~\ref{sec:early_late}, we discuss the tension between early-Universe and late-Universe probes of $H_{0}$.  We summarize our findings in Section~\ref{sec:summary}.  Throughout this paper, all magnitudes given are on the AB system.  All parameter constraints given are medians and 16th and 84th percentiles unless otherwise stated.

\section{Time-Delay Cosmography} \label{sec:tdcosmo}
In this section, we summarize the theoretical background of time-delay cosmography and how to infer $H_{0}$.  We refer readers to recent reviews \citep[e.g.,][]{treumarshall2016, suyu+2018} for more details.

When light rays from a background source are deflected by an intervening lensing mass, the light travel time from
the source to the observer depends on both their path length and the gravitational potential they traverse.  For a single lens plane, the excess time delay of an image at an
angular position $\bm{\theta} = (\theta_{1}, \theta_{2})$ with
corresponding source position $\bm{\beta} = (\beta_{1}, \beta_{2})$
relative to the case of no lensing is
\begin{equation} \label{eq:excess_td}
t(\bm{\theta}, \bm{\beta}) = \frac{\tdist}{c} \left[ \frac{(\bm{\theta} - \bm{\beta})^{2}}{2} - \psi(\bm{\theta}) \right],
\end{equation}
where $\tdist$ is the time-delay distance and
$\psi(\bm{\theta})$ is the lens potential.  The time-delay distance \citep{refsdal1964,schneider+1992,suyu+2010} is
defined as
\begin{equation} \label{eq:ddt}
\tdist \equiv (1+\zd) \frac{\Dd \Ds}{\Dds},
\end{equation}
where $\zd$ is the lens redshift, $\Dd$ is the angular diameter
distance to the lens, $\Ds$ is the angular diameter distance to the
source, and $\Dds$ is the angular diameter distance between the lens
and the source.  $\tdist$ has units of distance and is inversely proportional to $H_{0}$, with weak dependence on other cosmological parameters.

If the alignment between the background source and the foreground lens is close enough, multiple images of the same background source are formed.  Light rays reaching the observer will have different excess time delays depending on which image they are observed at.  The time delay between two images of such a lens, $\Delta t_{ij}$, is the difference of their excess time delays,
\begin{equation} \label{eq:td}
\Delta t_{ij} = \frac{\tdist}{c} \left[ \frac{(\bm{\theta}_{i} - \bm{\beta})^{2}}{2} - \psi(\bm{\theta}_{i}) - \frac{(\bm{\theta}_{j} - \bm{\beta})^{2}}{2} + \psi(\bm{\theta}_{j}) \right],
\end{equation}
where $\bm{\theta}_{i}$ and $\bm{\theta}_{j}$ are the positions of images $i$ and $j$, respectively, in the image plane.  If the source is variable on short timescales, it is possible to monitor the fluxes of the images and measure the time delay, $\Delta t_{ij}$, between them \citep[e.g.,][]{vanderriest+1989,schechter+1997,fassnacht+1999,fassnacht+2002,kochanek+2006,courbin+2011}.  The lens potentials at the image positions, $\psi(\bm{\theta}_{i})$ and $\psi(\bm{\theta}_{j})$, as well as at the source position, $\bm{\beta}$, can be determined from a mass model of the system.  With a measurement of $\Delta t_{ij}$ and an accurate lens model to determine $\psi(\bm{\theta})$, it is possible to determine $\tdist$.  By further assuming a cosmological model, $\tdist$ can be converted into an inference on $H_{0}$.

If there are multiple lenses at different redshifts between the source and the observer, the observed time delays depend on combinations of the angular diameter distances among the observer, the multiple lens planes, and the source.  In this case, the image positions are described by the multiplane lens equation \citep[e.g.,][]{blandfordnarayan1986,kovner1987,schneider+1992,petters+2001,collettauger2014,mccully+2014}, and the observed time delays are no longer proportional to a single unique time-delay distance.  However, it is often the case that the mass in a single lens plane dominates the lensing effect, and the observed time delays are primarily sensitive to the time-delay distance (Equation~\ref{eq:ddt}) with the deflector redshift as that of the primary lens plane.  This is the case for all of the lenses in the H0LiCOW sample (Section~\ref{subsec:sample}).  The results for any individual system can thus be interpreted as a constraint on $D_{\Delta t} (z_{\mathrm{d}},z_{\mathrm{s}})$, which we refer to as the ``effective time-delay distance".  Hereafter, $\tdist$ refers to the effective time-delay distance (for applicable systems) unless otherwise indicated.

In addition to mass that is explicitly included in the lens model, all other mass along the LOS between the observer and the source contributes to the lens potential that the light rays traverse.  This causes additional focusing and defocusing of the rays and can affect the observed time delays \citep[e.g.,][]{seljak1994}.  If unaccounted for, this can lead to biased inferences of $\tdist$.  If the effects of the perturbing LOS masses are small, they can be approximated by an external convergence in the lens plane $\kext$ \citep[e.g.,][]{keeton2003,mccully+2014}.  The true $\tdist$ is related to the time-delay distance inferred from the lens model and measured time delays, $D_{\Delta t}^{\mathrm{model}}$, by the relation

 \begin{equation} \label{eq:ddtkappa}
\tdist = \frac{\tdistmod}{1-\kext}.
\end{equation}
$\kext$ is defined such that its average value across the sky is zero.  In principle, if lenses are randomly distributed, the effect of $\kext$ should average out over a sufficiently large sample.  However, the cross section for strong lensing scales as $\sigma^{4}$, where $\sigma$ is the velocity dispersion of the lens galaxy.  As a result, lenses are biased toward the most massive galaxies, which are known to cluster \citep[e.g.,][]{dressler1980}.  Indeed, lens galaxies generally lie in overdense environments and lines of sight relative to typical fields \citep[e.g.,][]{treu+2009,fassnacht+2011,wong+2018}, meaning that $\kext$ will lead to a bias on $\tdist$ and needs to be corrected for.  $\kext$ cannot, in general, be constrained from the lens model due to the mass-sheet degeneracy \citep{falco+1985,gorenstein+1988,saha2000}, in which the addition of a uniform mass sheet associated with a rescaling of the mass normalization of the strong lens galaxy and the coordinates in the source plane can modify the product of the time delays and $H_{0}$ but leave other observables unchanged.  $\kext$ must instead be estimated through other methods, such as studies of the lens environment or the use of lens stellar kinematics (as noted in the previous section).

With kinematic information on the lens galaxy, it is possible to determine the angular diameter distance to the lens, $\Dd$, independent of $\kext$ \citep{paraficzhjorth2009,jee+2015}.  Although the constraints from $\Dd$ are generally weaker than those from $\tdist$, it can break degeneracies among cosmological parameters, particularly for models beyond flat $\Lambda$CDM.  In particular, it can break the degeneracy between curvature ($\Ok$) and the time-varying equation of state parameter of dark energy ($w$) \citep{jee+2016}.  The combination of lensing, time delays, and lens kinematic data thus provides a joint constraint on $\tdist$ and $\Dd$ in cases of single strong-lensing planes \citep[see e.g.,][H0LiCOW IX for more details]{birrer+2016,chen+2019}.  These constraints on lensing distances, together with the redshifts of the lenses and sources, then allow us to infer cosmological parameter values for a given cosmological model.

\section{Overview of the H0LiCOW Analysis} \label{sec:analysis}
In this section, we provide a brief summary of the H0LiCOW analysis, including the sample of lenses analyzed to date (Section~\ref{subsec:sample}), as well as the various components that go into determining cosmological constraints from each lens.

\subsection{Lens Sample} \label{subsec:sample}
\begin{figure}
\centering
\subfloat[\blens]{
  \includegraphics[width=40mm]{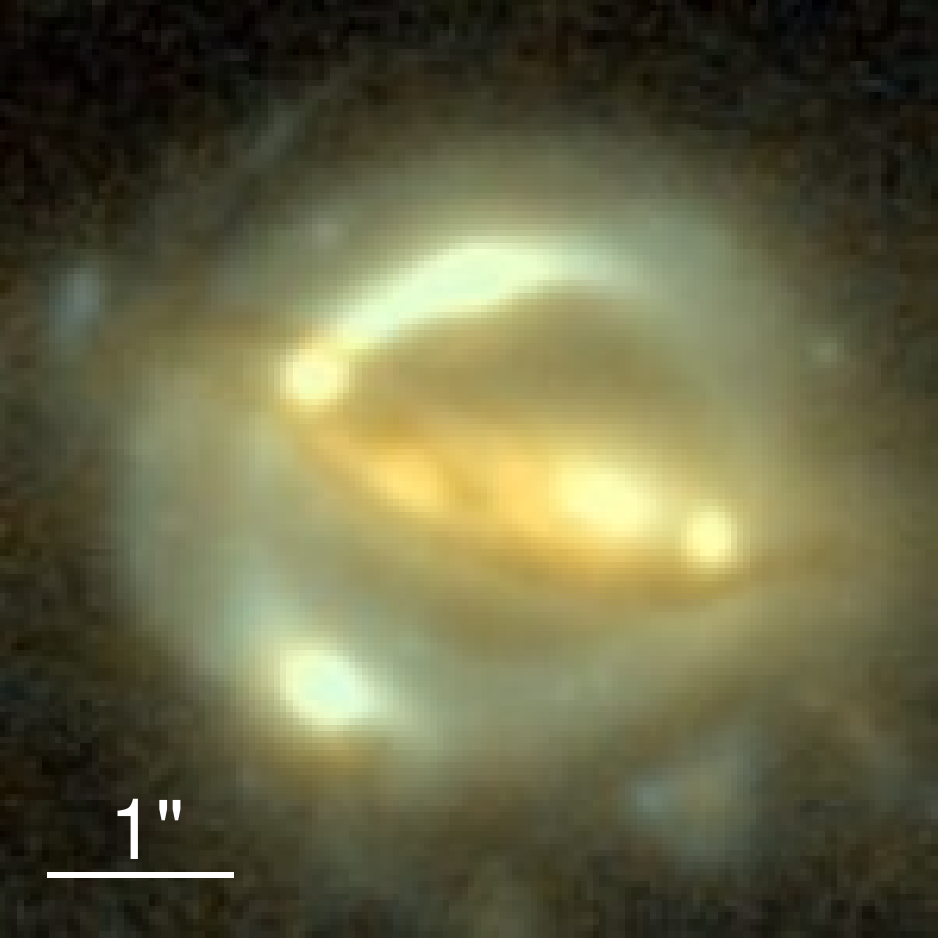}
}
\subfloat[\rxjlens]{
  \includegraphics[width=40mm]{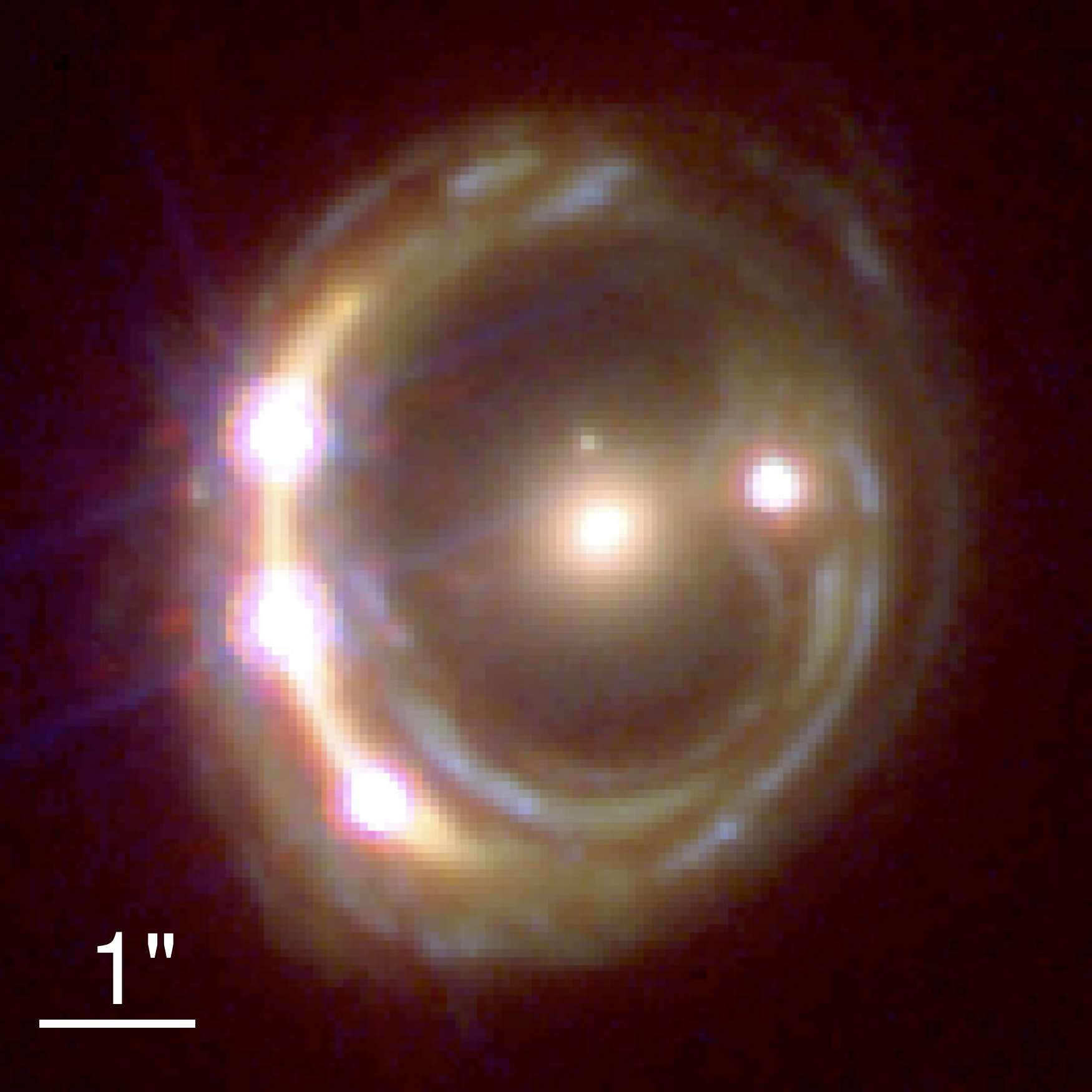}
}
\hspace{0mm}
\subfloat[\hequad]{
  \includegraphics[width=40mm]{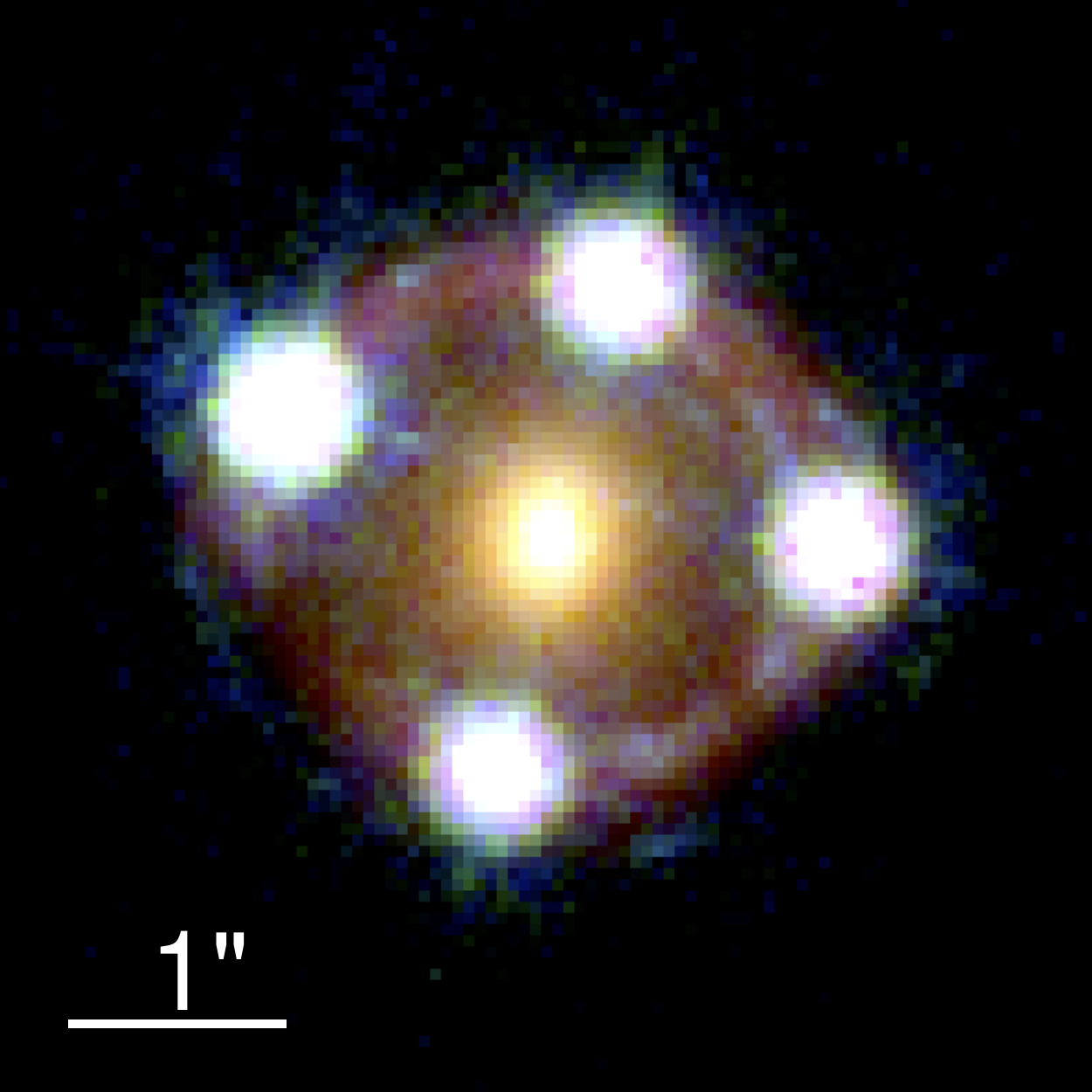}
}
\subfloat[\sdsslens]{
  \includegraphics[width=40mm]{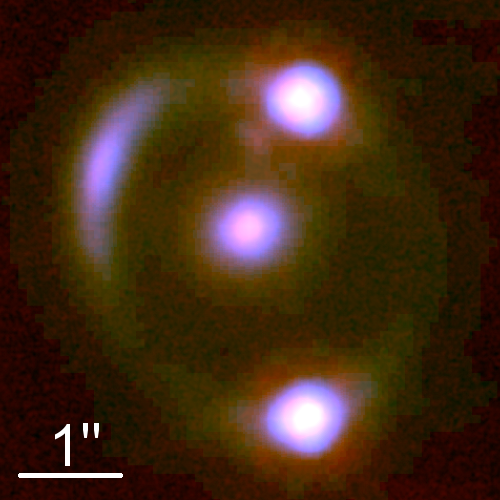}
}
\hspace{0mm}
\subfloat[\wfilens]{
  \includegraphics[width=40mm]{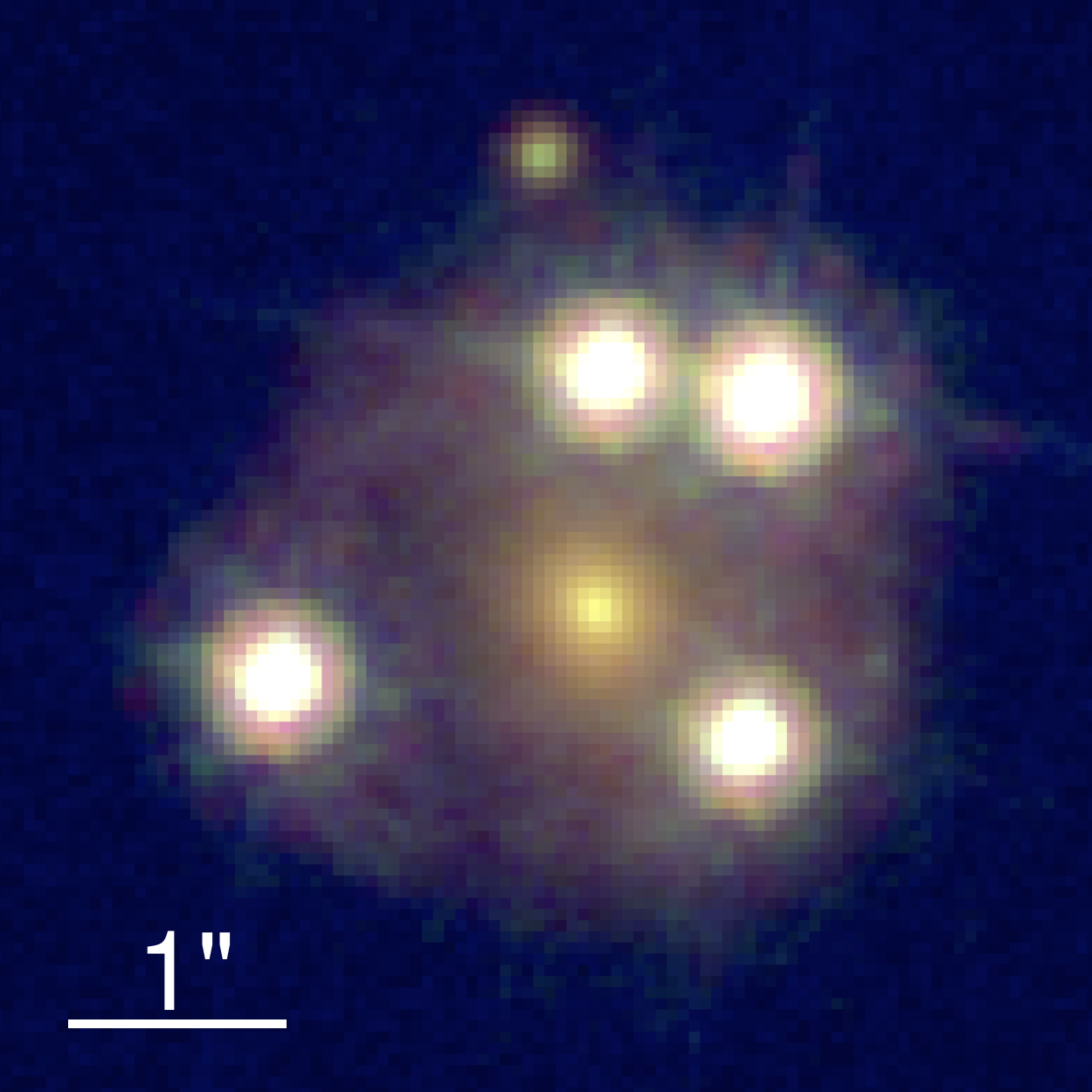}
}
\subfloat[\pglens]{
  \includegraphics[width=40mm]{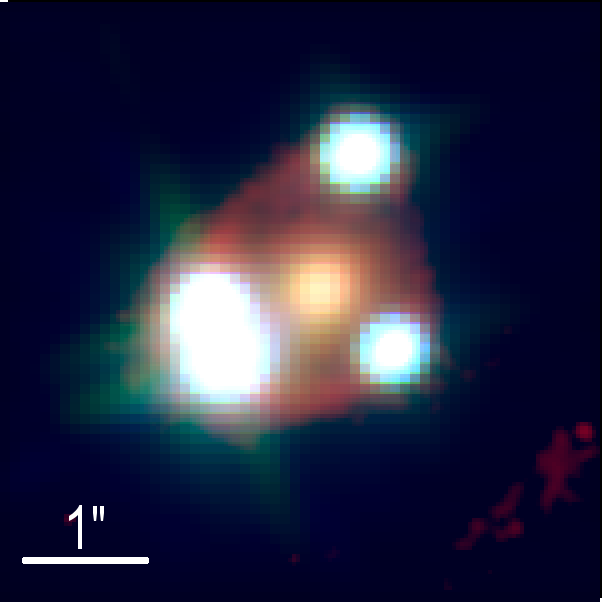}
}

\caption{Multicolor images of the six lensed quasars used in our analysis.  The images are created using two or three imaging bands in the optical and near-infrared from {\it HST} and/or ground-based AO data.  North is up and east is to the left.  Images for \blens, \rxjlens, \hequad, and \wfilens~are from H0LiCOW I.}
\label{fig:sample}
\end{figure}

Our sample of strongly-lensed quasars comprises six systems analyzed to date by H0LiCOW and collaborators.  The six lenses are listed in Table~\ref{tab:lenses}, and we show multicolor high-resolution images of them in Figure~\ref{fig:sample}.  Each of these systems have been modeled using constraints from high-resolution {\it HST} and/or ground-based AO imaging data, time-delay measurements from COSMOGRAIL and \citet{fassnacht+2002}, and kinematics from ground-based spectroscopy. In addition, we have constrained $\kext$ in these systems from a wide-field imaging and spectroscopic campaign, as detailed in Section~\ref{subsec:kext}.

The original H0LiCOW sample (see H0LiCOW I) consists of five lenses, and it was later decided to expand the sample to include four additional systems with {\it HST} imaging (PID:14254, PI: T. Treu), placing an emphasis on double-image lens systems (doubles), which yield fewer constraints per system but are more abundant on the sky.  Of the five doubles (one from the original sample plus the additional four), \sdsslens~was analyzed first (H0LiCOW IX) because part of the quasar's host galaxy is quadruply-imaged, providing additional constraints for lens modeling.  \pglens~was observed with AO imaging from Keck/NIRC2 as part of the Strong lensing at High Angular Resolution Program (SHARP; Fassnacht et al. in preparation) and was incorporated into the H0LICOW sample later \citep{chen+2019}.

The current sample of six systems used in this work include the four quadruple-image lenses (quads) from the original sample, plus \sdsslens~and \pglens.  These six lenses span a range of lens and source redshifts, as well as a range of image configurations (e.g., double, cross, fold, cusp).  Having a sample that spanned a range in these parameters was a consideration in the selection of which lenses to analyze first, as there may be systematics that depend on such factors, and we want to account for them in our analysis \citep[see][who attempt to address these issues based on simulated data]{ding+2018}.

\renewcommand*\arraystretch{1.5}
\begin{table*}
\caption{Lenses in the H0LiCOW sample used in this paper. \label{tab:lenses}}
\begin{minipage}{\linewidth}
\centering
\begin{tabular}{llllll}
\hline
Lens name &
$\alpha$ (J2000) &
$\delta$ (J2000) &
$z_{\mathrm{d}}$ &
$z_{\mathrm{s}}$ &
{\it HST} / AO data
\\
\hline
\blens$^{a}$ &
16:09:13.96 &
$+$65:32:29.0 &
0.6304$^{a}$ &
1.394$^{b}$ &
{\it HST}
\\
\rxjlens$^{c}$ &
11:31:51.6 &
$-$12:31:57.0 &
0.295$^{c}$ &
0.654$^{d}$ &
{\it HST} + AO
\\
\hequad$^{e}$ &
04:38:14.9 &
$-$12:17:14.4 &
0.4546$^{f,g}$ &
1.693$^{h}$ &
{\it HST} + AO
\\
\sdsslens$^{i}$ &
12:06:29.65 &
$+$43:32:17.6 &
0.745$^{j}$ &
1.789$^{i}$ &
{\it HST}
\\
\wfilens$^{k}$ &
20:33:41.9 &
$-$47:23:43.4 &
0.6575$^{l}$ &
1.662$^{h}$ &
{\it HST}
\\
\pglens$^{m}$ &
11:18:16.899 &
$+$7:45:58.502 &
0.311$^{n}$ &
1.722$^{m}$ &
{\it HST} + AO
\\
\hline
\end{tabular}
\\
\begin{tablenotes}
\item[$a$] \citet{myers+1995};
\item[$b$] \citet{fassnacht+1996};
\item[$c$] \citet{sluse+2003};
\item[$d$] \citet{sluse+2007};
\item[$e$] \citet{wisotzki+2002};
\item[$f$] \citet{morgan+2005};
\item[$g$] \citet{eigenbrod+2006};
\item[$h$] \citet{sluse+2012};
\item[$i$] \citet{oguri+2005};
\item[$j$] \citet{agnello+2016};
\item[$k$] \citet{morgan+2004};
\item[$l$] \citet{sluse+2019};
\item[$m$] \citet{weymann+1980};
\item[$n$] \citet{tonry1998}
\end{tablenotes}
\end{minipage}
\end{table*}
\renewcommand*\arraystretch{1.0}

\subsection{Time Delay Measurement} \label{subsec:td}
Out of the six lenses of the H0LICOW sample, all except for \blens~have been monitored in optical by the COSMOGRAIL collaboration from several facilities with 1m and 2m-size  telescopes. Several seasons of monitoring are needed in order to disentangle the variations due to microlensing in which brightening or dimming of the quasar images by stars in the lens galaxy can mimic intrinsic features in the light curves.
  
From the monitoring data, COSMOGRAIL measures time delays using numerical curve-shifting techniques, which fit a function to the light curve of each quasar image and find the time shifts that minimize the differences among them \citep{tewes+2013a,bonvin+2019}.  These techniques are made publicly available as a {\sc python} package named \pycs \footnote{Available at http://www.cosmograil.org}, which also provides tools to estimate the time delays uncertainties in the presence of microlensing. 
The package was tested on simulated light curves reproducing the COSMOGRAIL data with similar sampling and photometric noise in a blind time delay challenge \citep{liao+2015}. \cite{bonvin+2016} demonstrated the robustness of the \pycs curve-shifting techniques by recovering the time delays at a precision of $ \sim $ 3\% on average with negligible systematic bias. 

\cite{tewes+2013b} applied these techniques to \rxjlens~ and measured the longest time delay to 1.5\% precision (1$\sigma$).
The time delay of \sdsslens~was also measured with \pycs; \cite{eulaers+2013} obtained a time delay between the two multiple images of $\Delta t_{AB} = 111.3 \pm 3$ days, with image A leading image B. \cite{birrer+2019} re-analyzed the same monitoring data with updated and independent curve-shifting techniques and confirmed this result. For \hequad, the latest time delay measurement was obtained with the 13 year-long light curves of the COSMOGRAIL program at 6.5\% precision on the longest time delay \citep{bonvin+2017}. 

Recently, \cite{courbin+2018} demonstrated that a high-cadence and high signal-to-noise (S/N) monitoring campaign can also disentangle the microlensing variability from the intrinsic variability signal by catching small variations of the quasar that happen on timescales much shorter than the typical microlensing variability. It is therefore possible to disentangle the intrinsic signal of the quasar from the microlensing signal in a single season. High-cadence data were used for \wfilens~and \pglens~to measure time delays at a few percent precision in one season. These results are in agreement with the time delays measured from decade-long COSMOGRAIL light curves and are combined in the final estimate \citep{bonvin+2018, bonvin+2019}. 

The remaining lens of the sample, \blens~ was monitored by \cite{fassnacht+1999, fassnacht+2002} with radio observations from the Very Large Array over three seasons. All three independent time delays between the multiple images were measured to a precision of a few percent.

A complicating factor in converting the observed time delays to a cosmological constraint is the so-called ``microlensing time-delay" effect \citep{tiekochanek2018}. The estimation of this effect is based on the lamp-post model, which predicts delayed emission across the quasar accretion disk from a central driving source. Different regions of the disk can then be magnified by the microlenses differently in each of the multiple images. This reweighting of the delayed emission across the accretion disk could lead to a change in the measured time delay. As the microlensing changes with time, this could lead to a variation in the measured time delays from season to season.  There is no evidence of this effect based on our current data, so our main cosmological results do not depend on it.  Nonetheless, we quantify this factor for different speculative models \citep{bonvin+2018,bonvin+2019,chen+2018} in the latest H0LiCOW lens models \citep{birrer+2019,chen+2019,rusu+2019}.

\subsection{Lens Modeling} \label{subsec:lensmodel}
The primary lens modeling code used to model the majority of the H0LiCOW lenses is {\sc Glee} \citep{suyuhalkola2010,suyu+2012}, although \sdsslens~(H0LiCOW IX) is analyzed using the {\sc Lenstronomy} code \citep{birrer+2015,birreramara2018}.  Both codes model the lens galaxy light as parameterized profiles and fit the lensed quasar image positions and surface brightness distribution of the quasar host galaxy.  The primary difference between the codes is that {\sc Glee} reconstructs the source on a pixelized grid with regularization, whereas {\sc Lenstronomy} describes the source as a parameterized profile with additional shapelet functions \citep{refregier2003,birrer+2015}.  The use of two independent codes is meant to provide a check on lens modeling code systematics.  This would ideally require both codes to be tested on the same system, as will be done for future lens analyses (Shajib et al. in preparation; Y{\i}ld{\i}r{\i}m et al. in preparation).

We use two main parameterizations of the lens galaxy in our models: a singular elliptical power-law model, and a composite model consisting of a baryonic component linked to the stellar light distribution plus an elliptical NFW \citep{navarro+1996} halo representing the dark matter component.  For \blens, which shows two interacting lens galaxies, rather than using the two main parameterizations, we started from the power-law model and performed a pixelated lens potential reconstruction to allow for flexibility, finding small ($\sim2\%$) potential corrections and thus validating the use of power-law model.  Galaxies along the LOS that are deemed to be significant perturbers are included in the model (Section~\ref{subsec:kext}) through the full multiplane lens equation, and we also include an external shear in the main lens plane.  When available, we use the measured velocity dispersions of the lens galaxy and significant perturbers as additional constraints.

To account for systematic effects arising from modeling choices in areas such as the lens parameterization, the source reconstruction, the weighting of the pixels in the image plane, etc., we run multiple models where we vary these choices and combine them in our final inference.  In our initial analyses of the first three H0LiCOW lenses \citep[][H0LiCOW IV]{suyu+2010,suyu+2013,suyu+2014}, we marginalize over these (discrete) modeling choices in deriving the posterior probability density function (PDF) of $\tdist$.  For models that fit equally well to the data within their modeling uncertainties, we conservatively add their posterior distributions of $\tdist$ with equal weight, given uniform prior on these modeling choices.  For subsequent analyses, including the reanalysis of \rxjlens~and \hequad~\citep{chen+2019}, we weight by the Bayesian Information Criterion (BIC), following the procedure described in H0LiCOW IX.

We have measurements of the velocity dispersions of the lens galaxies in our sample from high-resolution spectroscopy, which are used to mitigate degeneracies in the mass modeling.  The velocity dispersion can be combined with the lensing constraints to estimate the angular diameter distancesto the lens (see Section~\ref{sec:tdcosmo}), in the systems (\blens, \rxjlens, \sdsslens\ and \pglens) that could be well-modeled by single lens plane (i.e., without multi lens plane modeling).

\subsection{LOS Structure and External Convergence} \label{subsec:kext}
In accounting for the effects of LOS structure, there are two primary types of perturbations that need to be dealt with.  The first is the effect of structures that affect the lens potential significantly enough that they cannot be approximated by their tidal perturbations, but must instead be included explicitly in the lens model (Section~\ref{subsec:lensmodel}).  The second is the combined effect of all other LOS structures, which can be approximated by a $\kext$ term.  Quantifying and accounting for both types of perturbers requires spectroscopic and photometric data on LOS objects projected nearby the lens.

The effect of a LOS perturber on the lens potential can be quantified by the ``flexion shift" \citep{mccully+2014}.  In general, objects have a larger flexion shift if they are more massive, projected more closely to the lens, and are either at the redshift of the lens or at a lower redshift, as opposed to a higher redshift \citep{mccully+2017}.  As a result, we focus our spectroscopic campaign on bright galaxies projected close to the lens and include them in our lens model if their calculated flexion shift is large. When their velocity dispersions are measurable from our data, we use them to set a prior on their Einstein radii during the modeling procedure (e.g., H0LiCOW X).  Alternatively, a prior on the Einstein radius is set from scaling relations with luminosity (e.g., H0LiCOW IX). An overview of our spectroscopic campaigns is provided in H0LiCOW I for the majority of our sample of lenses, and in H0LiCOW IX for \sdsslens . For \pglens, where we have not conducted spectroscopic follow-up, we use the compilation of redshifts presented by \citet{wilson+2016}, which was also useful in providing additional redshifts for some of the other lens fields.
Our spectroscopic data provide accurate redshifts for hundreds of bright galaxies as far as $\sim10\arcmin$ away from the lens systems, allowing us to further quantify the properties of larger structures such as galaxy groups and clusters. Since the analysis of \hequad~(H0LiCOW II), we use an adaptive group-finding algorithm that uses spectroscopic data to identify peaks in redshift space and refines group membership based on proximity between potential group galaxies (e.g., LOS velocity dispersion and centroid; see H0LiCOW X for details). In cases where there are potentially significant effects from such structures, we run systematics tests (Section~\ref{subsec:lensmodel}) where we include these structures as spherical NFW halos in our models.

To correct for the statistical effect of $\kext$ due to LOS structure, we use a (weighted) galaxy number counts technique \citep[e.g.,][H0LiCOW III, IX]{greene+2013,suyu+2013}. We count galaxies projected within a fixed distance of the lens and above some flux threshold, weighted by various quantities such as external shear, projected distance, stellar mass, redshift, etc. (see H0LiCOW III, XII for details). We then compare these number counts to those measured analogously along random lines of sight in a control survey to determine the relative over-/under-density of the lens field. Finally, we use the Millennium simulation \citep{springel+2005} to identify lines of sight that have a similar relative number count densities and build a PDF of $\kext$ determined from ray-tracing \citep{hilbert+2009}, which we apply in post-processing.  Alongside our spectroscopic campaign, we have conducted our own multi-band, wide-field imaging campaign to gather data of sufficient depth and spatial coverage to enable this analysis. 
Our data typically consist of multi-band ultraviolet to infrared observations in good seeing conditions, which we use to perform a galaxy-star classification and to measure physical quantities such as redshifts and stellar masses for the galaxies projected within $<120\arcsec$ of the lenses, and down to $i\sim23-24$ mag. Exceptions are \blens~and \rxjlens, where we used single-band {\it HST} data within $<45\arcsec$ \citep{fassnacht+2011}, and \pglens, where we coadded multiple exposures of the data used to measure the time delays. With the exception of the first two lenses we analyzed (\blens~and \rxjlens) where we used archival {\it HST} data as control fields, we employed the larger-scale CFHTLenS \citep{heymans2012}.

Our technique has evolved over the years such that for \blens~we only employed unweighted number counts to constrain $\kext$, whereas for the remaining lenses we also used constraints from the inferred external shear values of the lens models.\footnote{An exception is \sdsslens, where the use of external shear has a negligible effect on $\kext$.} Since \hequad~(H0LiCOW III), we have further used combinations of weighted counts to tighten the $\kext$ PDF, which for the three latest lenses have included weighted number counts measured in multiple apertures.

In future work, we plan to return to previous lenses in order to enforce consistency of technique, as well as to further refine our technique by better accounting for lens-lens coupling between the primary lens and LOS structures in the convergence maps from \citet{hilbert+2009}, employing other cosmological simulations with different assumed cosmology, and also using new techniques which move away from the statistical approach and have the potential to significantly tighten the $\kext$ PDF \citep[e.g.,][]{mccully+2017}. In addition, in some of the H0LiCOW systems, we have independently constrained the external convergence using weak lensing (\citeauthor{tihhonova+2018} \citeyear{tihhonova+2018}, hereafter H0LiCOW VIII, and Tihhonova et al.~submitted, hereafter H0LiCOW XI). The external convergence determined through weak lensing is consistent with our weighted number counts calculation.

\subsection{Joint Inference} \label{subsec:inference}
For our analysis, we make use of multiple datasets, denoted by $\bm{d_{\mathrm{img}}}$ for the
\hst~and (if available) AO imaging data, $\bm{\Delta t}$ for the time delays, $\bm{\sigma}$ for the
velocity dispersion of the lens galaxy, and $\bm{d_{\mathrm{LOS}}}$ for
the properties of the LOS mass distribution determined from our
photometric and spectroscopic data. We want to obtain the posterior
PDF of the model parameters
$\bm{\xi}$ given the data, $P(\bm{\xi} | \bm{d_{\rm img}, \Delta t, \sigma,
d_{\mathrm{LOS}},A})$.  The vector~$\bm{\xi}$ includes the lens model
parameters~$\bm{\nu}$,
the cosmological parameters $\bm{\pi}$,
and nuisance parameters representing the external convergence ($\kext$) and anisotropy radius for the lens stellar velocity ellipsoid ($r_{\rm ani}$).
${\bm A}$ denotes a discrete set of assumptions about the form of the
model, including the choices we have to make about the data modeling region, the set-up of the source reconstruction grid, the treatment of
the different deflector mass distributions, etc.  In general, ${\bm A}$ cannot be fully captured by continuous parameters.
From Bayes' theorem, we have that
\begin{eqnarray} \label{eq:pdf}
&& P(\bm{\xi} | \bm{d_{\rm img}, \Delta t, \sigma, d_{\mathrm{LOS}}, A}) \nonumber \\
        &\propto& P(\bm{d_{\rm img}, \Delta t, \sigma, d_{\mathrm{LOS}}} | \bm{\xi,A}) P(\bm{\xi} | \bm{A}),
\end{eqnarray}
where ${P(\bm{d_{\rm img}, \Delta t, \sigma, d_{\mathrm{LOS}}}} | \bm{\xi,
A})$ is the joint likelihood function and $P(\bm{\xi} | \bm{A})$ is the
prior PDF for the parameters given our assumptions.  Since the data sets
are independent, the likelihood can be separated,
\begin{eqnarray} \label{eq:pdf_sep}
P(\bm{d_{\rm img}, \Delta t, \sigma, d_{\mathrm{LOS}}} | \bm{\xi, A}) &=& P(\bm{d_{\rm img}} | \bm{\xi, A}) \nonumber \\
  &&  \times P(\bm{\Delta t} | \bm{\xi, A}) \nonumber \\
  &&  \times P(\bm{\sigma} | \bm{\xi, A})  \nonumber \\
  &&  \times P(\bm{d_{\mathrm{LOS}}} | \bm{\xi, A}).
\end{eqnarray}
We can calculate the individual likelihoods separately and combine them as in \eref{eq:pdf_sep} to get the final posterior PDF for a given set of assumptions.

For any given lens model, we can vary the content of $\bm{A}$ and repeat the inference
of ${\bm \xi}$.  This can be important for checking
the impact of various modeling choices and assumptions, but
leaves us with the question of how to combine the results.  Depending on the lens, we either combine the models with equal weight, or we can use the Bayesian Information Criterion (BIC) to weight the different models in our final inference \citep[e.g., H0LiCOW IX, XII,][]{chen+2019}.  This effectively combines our various assumptions $\bm{A}$ using the BIC so that we obtain $P(\bm{\xi} | \bm{d_{\rm img}, \Delta t, \sigma, d_{\mathrm{LOS}}})$.  We can further marginalize over the non-cosmological parameters ($\bm{\nu}$, $\kext$, $r_{\rm ani}$) and obtain the posterior probability distribution of the cosmological parameters $\bm{\pi}$:
\begin{eqnarray} \label{Pcosmo}
  & & P(\bm{\pi} | \bm{d_{\rm img}, \Delta t, \sigma, d_{\mathrm{LOS}}}) \nonumber \\
  & = &  \int {\rm d}{\bm \nu}\, {\rm d}\kext\, {\rm d}r_{\rm ani}P(\bm{\xi} | \bm{d_{\rm img}, \Delta t, \sigma, d_{\mathrm{LOS}}}).
\end{eqnarray}

In the lens modeling of systems with a single strong lens plane, the parameters associated with cosmology that enter directly into the lens modeling are the two lensing distances $\tdist$ and $\Dd$.  In the lens modeling of systems with multiple strong lens planes, we actually vary $H_{0}$, keeping other parameters fixed at $w = -1$, $\Om= 0.3$, and $\OL = 0.7$.  This assumes a fixed curvature of the expansion history of the Universe, but not the absolute scale (represented by $H_{0}$ or $\tdist$).  This is done because there is not a unique $\tdist$ when accounting for multiple lens planes, but we convert this to an ``effective" $\tdist$ that is insensitive to assumptions of the cosmological model. Specifically, given the lens/quasar redshifts and $\bm{\pi}$ (i.e., $H_{0}$ and the other fixed cosmological parameters), we can compute the effective time-delay distance $\tdist(\bm{\pi}, \zd, \zs)$ to obtain the posterior probability distribution of $\tdist$, $P(\tdist |  \bm{d_{\rm img}, \Delta t, \sigma, d_{\mathrm{LOS}}})$.  In summary, the single-lens plane models yield a joint constraints on $\tdist$ and $\Dd$, whereas multi-lens plane models yield a constraint on the effective $\tdist$.

\subsection{Blind Analysis} \label{subsec:blind}
After the development of the lens modeling and analysis methods that were first applied to \blens, the subsequent five lenses in H0LiCOW are analyzed blindly with respect to the cosmological quantities of interest (i.e., $\tdist$, $\Dd$, $H_{0}$).  Throughout the analyses, these values are blinded by subtracting the median of their PDF from the distribution.  This allows us to view the shape of the distribution, their relative shifts, as well as covariances with other model parameters without ever seeing the absolute value.  This is done to prevent confirmation bias and to remove the tendency for experimenters to stop analyzing systematic errors once they have achieved a result that agrees with a prior ``expected" value.  When the analysis of a particular H0LiCOW lens is finished and all team members have agreed to show the results, we unblind the relevant parameters and publish the result with no further changes to the analysis.

\subsection{Distance Constraints and Error Budget for the Sample} \label{subsec:tddist}
We list the $\tdist$ and (when available) $\Dd$ constraints for each individual lens in Table~\ref{tab:dt}, along with corresponding references.  All distances listed here are used in our cosmological inference.  Specifically, for \blens, we use the analytic fit of $P(\tdist)$ given in \citet{suyu+2010} and of $P(\Dd)$ given in \citet{jee+2019}, and multiply these two PDFs since these two distances are uncorrelated for this system.  For \hequad, \rxjlens\ and \pglens, we use the resulting Monte Carlo Markov chains (MCMC) of $\tdist$ and (if available) $\Dd$ from the analysis of \citet{chen+2019}, which includes the previous \hst\ constraints from \citet{suyu+2014} and H0LiCOW IV, as well as new AO data.  For \sdsslens, we use the resulting MCMC chain of $\tdist$ and $\Dd$ from H0LiCOW IX.  For \wfilens, we use the resulting MCMC chain of $\tdist$ from H0LiCOW XII.  We use a kernel density estimator to compute $P(\tdist,\Dd)$ or $P(\tdist)$ from the chains, allowing us to account for any correlations between $\tdist$ and $\Dd$ in $P(\tdist,\Dd)$.

\renewcommand*\arraystretch{1.5}
\begin{table*}
\caption{$\tdist$ and $\Dd$ constraints for H0LiCOW lenses. \label{tab:dt}}
\begin{minipage}{\linewidth}
\centering
\begin{tabular}{l|llll}
\hline
Lens name &
$\tdist$ (Mpc) &
$\Dd$ (Mpc) &
Blind analysis &
References
\\
\hline
\blens &
$5156_{-236}^{+296}$ &
$1228_{-151}^{+177}$&
no &
\citet{suyu+2010,jee+2019}
\\
\rxjlens &
$2096^{+98}_{-83}$ &
$804^{+141}_{-112}$ &
yes$^{a}$ &
\citet{suyu+2014,chen+2019}
\\
\hequad &
$2707_{-168}^{+183}$ &
--- &
yes &
\citet{wong+2017,chen+2019}
\\
\sdsslens &
$5769_{-471}^{+589}$ &
$1805_{-398}^{+555}$ &
yes &
\citet{birrer+2019}$^{b}$
\\
\wfilens &
$4784_{-248}^{+399}$ &
--- &
yes &
\citet{rusu+2019}
\\
\pglens &
$1470_{-127}^{+137}$ &
$697_{-144}^{+186}$ &
yes &
\citet{chen+2019}
\\
\hline
\end{tabular}
\\
{\footnotesize Reported values are medians, with errors corresponding to the 16th and 84th percentiles.  Values for \blens~are calculated from a skewed lognormal function fit to the posterior distributions.  Values for \rxjlens, \hequad, \sdsslens, \wfilens, and \pglens~are calculated from the MCMC chains of the posterior distributions of the distances.
For \rxjlens, \sdsslens\ and \pglens, we use the joint $P(\tdist, \Dd)$ distributions for cosmographic inferences (see Section \ref{subsec:tddist} for details) to account for correlations between $\tdist$ and $\Dd$. }
\begin{tablenotes}
\item[$a$] The initial {\it HST} analysis \citep{suyu+2013} was performed blindly, but the AO analysis \citep{chen+2019} was not.
\item[$b$] The values given here are updated values from those presented in \citet{birrer+2019}, which had a minor error in the calculation of the 16th and 84th percentiles.  The median values are unchanged, while the uncertainties have changed by $\sim3\%$.
\end{tablenotes}
\end{minipage}
\end{table*}
\renewcommand*\arraystretch{1.0}

We estimate the approximate $\tdist$ error budget for each of the lenses in Table~\ref{tab:err_budget}.  The contributions from the time delay measurement and LOS calculation are based on a Gaussian approximation.  The remainder of the uncertainty is attributed to the lens model and other sources, which are difficult to disentangle.  This breakdown shows that there is no single source of error that dominates the uncertainty from time-delay cosmography in general.  Rather, it depends on characteristics of each particular lens that can be effectively random \citep[modulo certain biases that make the lens more likely to be discovered, although such biases affect distance measurements at $\lesssim1\%$ level;][]{Collett+2016}, such as the image configuration (affects time delays and modeling), the mass/ellipticity of the lens galaxy (affects image separation/time delays, the lens model, and can be linked to the overdensity of the local environment), LOS structure (which is mostly uncorrelated outside of the local lens environment), and other factors.  Moving forward, efforts will have to be made to tackle all of these sources of error to improve constraints from the overall sample rather than focusing on a single factor.  Alternatively, with a large enough sample, one can pick out a small number of ``golden lenses" that have characteristics that make them likely to have small uncertainties from each of the contributing sources of error, although one would have to be careful about potential biases in culling such a subsample.

\renewcommand*\arraystretch{1.5}
\begin{table*}
\caption{Approximate $\tdist$ error budget for H0LiCOW lenses. \label{tab:err_budget}}
\begin{minipage}{\linewidth}
\centering
\begin{tabular}{l|cccccc}
\hline
Source of uncertainty &
\blens &
\rxjlens &
\hequad &
\sdsslens &
\wfilens &
\pglens
\\
\hline
Time delays &
1.7\% &
1.6\% &
5.3\% &
2.3\% &
2.9\% &
6.4\%
\\
Line-of-sight contribution &
6.4\% &
3.3\% &
2.8\% &
2.9\% &
5.7\% &
2.7\%
\\
Lens model and other sources &
3.0\% &
2.2\% &
2.5\% &
8.4\% &
2.2\% &
5.7\%
\\
\hline
Total &
5.1\% &
4.3\% &
6.5\% &
9.1\% &
6.7\% &
9.0\%
\\
\hline
\end{tabular}
\\
{\footnotesize Approximate $\tdist$ error budget for each of the six lenses in the H0LiCOW sample. The contributions from the time-delay measurement and LOS calculation are based on a Gaussian approximation to the PDFs of $\Delta t$ and $\kext$.  Specifically, the uncertainty is obtained by taking half-width of the 68\% credible interval of the corresponding PDF and dividing it by the median value.  The remainder of the uncertainty is attributed to the lens model and other sources, which are difficult to disentangle.  Nonetheless, they are estimated such that the total uncertainty on $\tdist$ (computed from the posterior PDF of $\tdist$ using the median value and half of the 68\% credible interval) is the sum of the different sources in quadrature.  An exception is \blens, where the uncertainty in $\kext$ is larger than the total uncertainty since the lens kinematic measurement excludes parts of the $P(\kext)$ distribution in the cosmological models considered in \citet{suyu+2010}.  For this case, we report the lens model uncertainty as that estimated from the power-law slope of the lens mass profile (which scales approximately linearly with $\tdist$).}
\end{minipage}
\end{table*}
\renewcommand*\arraystretch{1.0}

\section{Checking Consistency Among Lenses} \label{sec:combine_lenses}
We check that all our lenses can be combined without any loss
of consistency by comparing their $\tdist$
posteriors in the full cosmological parameter space and
measuring the degree to which they
overlap. We quantify the consistency by using the Bayes factor (or evidence ratio) $F$ in favor of a simultaneous fit of the lenses using a common set of cosmological parameters \citep[e.g.,][]{marshall+2006,suyu+2013,bonvin+2017}.  When comparing data
sets $\bf{d_1},...,\bf{d_n}$, we can either assume the
hypothesis $\rm{H^{global}}$ that they can be represented using a
common global set of cosmological parameters, or the hypothesis $\rm{H^{ind}}$ that
at least one data set is better represented using a different
set of cosmological parameters. We emphasize that the latter model would 
make sense if there is a systematic error present that leads to a vector offset in
the inferred cosmological parameters. Parameterizing this offset vector
with no additional information would take as many nuisance parameters as
there are dimensions in the cosmological parameter space; assigning uninformative
uniform priors to each of the offset components is equivalent to using
a complete set of independent cosmological parameters for the outlier dataset.

We can compute the Bayes factor between any two lenses,
\begin{align} \label{eq:f_pairwise}
 F_{ij} &= \frac{P({\bf d_i}, {\bf d_j} | \mathrm{H^{global})}}{P({\bf d_i} | \mathrm{H^{ind}}) P({\bf d_j} | \mathrm{H^{ind}})} \\
 &= \frac{\langle {L_i} {L_j} \rangle}{\langle {L_i} \rangle \langle {L_j}\rangle},
\end{align}
where $L_{i}$ and $L_{j}$ are the likelihoods of data sets $\mathrm{{\bf d_{i}}}$ and $\mathrm{{\bf d_{j}}}$, respectively.  If the six lenses have Bayes factors $F > 1$ for every possible pairwise combination, it means that the lenses are statistically consistent with each other and we can proceed to combine their constraints.

In Table \ref{tab:bayes}, we show that none of the 15 possible pairwise combinations of the six lens systems have a Bayes factor $F < 1$. The minimal Bayes factor is obtained for the pair \pglens~- \sdsslens\ with $F = 2.7$, still favoring the $\rm{H^{global}}$ hypothesis. We also test the hypothesis that one out of six systems is better represented in a different set of cosmological parameters than the five remaining lenses.  The minimal Bayes factor is obtained for \rxjlens~with $F=3.8$, again in favor of the $\rm{H^{global}}$ hypothesis, meaning that all lenses are a consistent realization of the same underlying set of cosmological parameters.  We conclude that none of the six lenses is in disagreement with the cosmological parameters inferred from the five other systems.  This is an important check of the consistency of our results.  If our uncertainties were underestimated, we would not necessarily expect all of our lenses to give statistically consistent results.

\renewcommand*\arraystretch{1.5}
\begin{table*}
\caption{Bayes factor for all pairs of lensed systems (top) and of every individual system relative to the five remaining systems (bottom).\label{tab:bayes}}
  \begin{minipage}{\linewidth}
  \centering
\begin{tabular}{l|cccccc}
\hline
\multicolumn{7}{c|}{Pairwise Bayes factors}        \\ \hline \hline
\multicolumn{1}{l|}{}             & \blens                & \rxjlens              & \hequad               & \sdsslens & \wfilens & \pglens \\ \hline
\multicolumn{1}{l|}{\blens}       & \multicolumn{1}{c}{---} & 3.4                   & 10.6                  & 9.1       & 11.2     & 3.3     \\
\multicolumn{1}{l|}{\rxjlens}     &                       & \multicolumn{1}{c}{---} & 5.3                   & 3.1       & 4.9      & 6.9     \\
\multicolumn{1}{l|}{\hequad}      &                       &                       & \multicolumn{1}{c}{---} & 7.8       & 9.3      & 3.9     \\
\multicolumn{1}{l|}{\sdsslens}    &                       &                       &                       & ---         & 7.9      & 2.7     \\
\multicolumn{1}{l|}{\wfilens}     &                       &                       &                       &           & ---        & 3.7     \\
\multicolumn{1}{l|}{\pglens}      &                       &                       &                       &           &          & ---       \\  
\hline
\multicolumn{7}{c}{Individual lens Bayes factors vs. rest of sample}                               \\ \hline \hline
\multicolumn{1}{l|}{}             & \blens                & \rxjlens              & \hequad               & \sdsslens & \wfilens & \pglens \\ \hline
\multicolumn{1}{l|}{Bayes Factor} & 11.1                  & 3.8                   & 11.0                  & 7.5       & 12.6     & 4.5   \\
\hline
\end{tabular}
\\
{\footnotesize The lowest Bayes factor is obtained for \pglens~and \sdsslens~with an evidence ratio of 2.7, still favoring the hypothesis that both distributions are two realizations of the same set of cosmological parameters.}
\end{minipage}
\end{table*}
\renewcommand*\arraystretch{1.0}

\section{Results of Cosmographic Analysis} \label{sec:cosmo}
We list the cosmological models considered in our analysis in Table~\ref{tab:cosmo_models}.  We distinguish between models where we use constraints from strong lenses alone (Sections~\ref{subsec:lcdm_lensonly} and~\ref{subsec:ext_lensonly}) from those in which we combine our constraints with other probes via importance sampling (Section~\ref{subsec:ext_comb}) or MCMC sampling (Section~\ref{subsec:snelenses}), even if the underlying cosmological model is the same.  For the analysis using strong lenses only, we adopt uniform priors on the cosmological parameters with ranges indicated in Table~\ref{tab:cosmo_models}.

\begin{table*}
  \caption{Description of the cosmological models considered in this work.} \label{tab:cosmo_models}
  \begin{minipage}{\linewidth}
  \centering
 \begin{tabular}{c|c|c}
 \hline
  Model name & Description & Priors\\
  \hline
  \multicolumn{3}{|c|}{Time-delay cosmography only} \\
  \hline
  U$\Lambda$CDM & Flat $\Lambda$CDM & $H_{0}$ uniform in [0, 150] km s$^{-1}$ Mpc$^{-1}$ \\ 
  & & $\Om=1-\OL$ \\
  & & $\Om$ uniform in [0.05, 0.5]\\
  \\
  Uo$\Lambda$CDM &  Open $\Lambda$CDM & $H_{0}$ uniform in [0, 150] km s$^{-1}$ Mpc$^{-1}$ \\
  & & $\Om$ uniform in [0.05, 5] \\
  & & $\Ok$ uniform in [$-$0.5, 0.5]\\
   & & $\OL=1-\Om-\Ok > 0$ \\ 
  \\
 U$w$CDM & Flat $w$CDM & $H_{0}$ uniform in [0,150] km s$^{-1}$ Mpc$^{-1}$  \\
  & & $\Om$ uniform in [0.05, 5] \\
  & & $w$ uniform in [$-$2.5, 0.5]\\
  & & $\Ode = 1 - \Om $ \\
    \\
 U$w_0w_a$CDM & Flat $w_0w_a$CDM & $H_{0}$ uniform in [0,150] km s$^{-1}$ Mpc$^{-1}$  \\
  & & $\Om$ uniform in [0.05, 5] \\
  & & $w_0$ uniform in [$-$2.5, 0.5]\\
   & & $w_a$ uniform in [$-$2, 2]\\
  & & $\Ode = 1 - \Om $ \\
  \hline
    \multicolumn{3}{|c|}{Time-delay cosmography combined with other probes} \\
  \hline
  $\Lambda$CDM & Flat $\Lambda$CDM & $\Om=1-\OL$ \\
  & & JLA/Pantheon for \{$H_{0}$, $\OL$\}\\
  \\
  o$\Lambda$CDM & Open $\Lambda$CDM & $\Om=1-\OL-\Ok > 0$ \\
  & & {\it Planck} (Section~\ref{subsubsec:olcdm_comb}) or JLA/Pantheon (Section~\ref{subsec:snelenses}) for \{$H_{0}$, $\OL$, $\Om$\}\\
  \\
  $w$CDM & Flat $w$CDM & $\Om=1-\Ode$ \\
  & & {\it Planck} (Section~\ref{subsubsec:wcdm_comb}) or JLA/Pantheon (Section~\ref{subsec:snelenses}) for \{$H_{0}$, $\Ode$, $w$\}\\
  \\
  $\mathrm{N_{eff}}\Lambda$CDM & Flat $\Lambda$CDM & {\it Planck} for \{$H_{0}$, $\OL$, $\mathrm{N_{eff}}$\}\\
  & Variable $\mathrm{N_{eff}}$ & \\
  \\
  $\mathrm{m}_{\nu}\Lambda$CDM & Flat $\Lambda$CDM & {\it Planck} for \{$H_{0}$, $\OL$, $\sum \mathrm{m}_{\nu}$\}\\
  & Variable $\sum\mathrm{m}_{\nu}$ & \\
  \\
  $\mathrm{N_{eff}m}_{\nu}\Lambda$CDM & Flat $\Lambda$CDM & {\it Planck} for \{$H_{0}$, $\OL$, $\mathrm{N_{eff}}$, $\sum \mathrm{m}_{\nu}$ \}\\
  & Variable  $\mathrm{N_{eff}}$ and $\sum\mathrm{m}_{\nu}$ & \\
  \\
  $w_{0}w_{a}$CDM & Flat $w_0$$w_a$CDM & {\it Planck} (Section~\ref{subsubsec:w0walcdm}) or JLA/Pantheon (Section~\ref{subsec:snelenses}) for \{$H_{0}$, $w_{0}$, $w_{a}$\} \\
  &  &  \\
  \\
  o$w$CDM & Open $w$CDM & $\Om=1-\Ode-\Ok > 0$ \\
  &  & JLA/Pantheon for \{$H_{0}$, $\Ok$, $\Ode$, $w$\} \\
  \\
  o$w_{0}w_{a}$CDM & Open $w_0$$w_a$CDM & $\Om=1-\Ode-\Ok > 0$ \\
  &  & JLA/Pantheon for \{$H_{0}$, $\Ok$, $\Ode$, $w_{0}$, $w_{a}$\}  \\
  \hline
 \end{tabular}
 \\
{\footnotesize {\it Planck} refers either to the constraints from {\it Planck} 2018 Data Release alone, or combined with CMBL or BAO.  JLA refers to the joint light-curve analysis of \citet{betoule+2014}.  Pantheon refers to the sample of \citet{scolnic+2018}.}
  \end{minipage}
\end{table*}

\subsection{Flat $\Lambda$CDM} \label{subsec:lcdm_lensonly}
\begin{figure*}
\includegraphics[width=\textwidth]{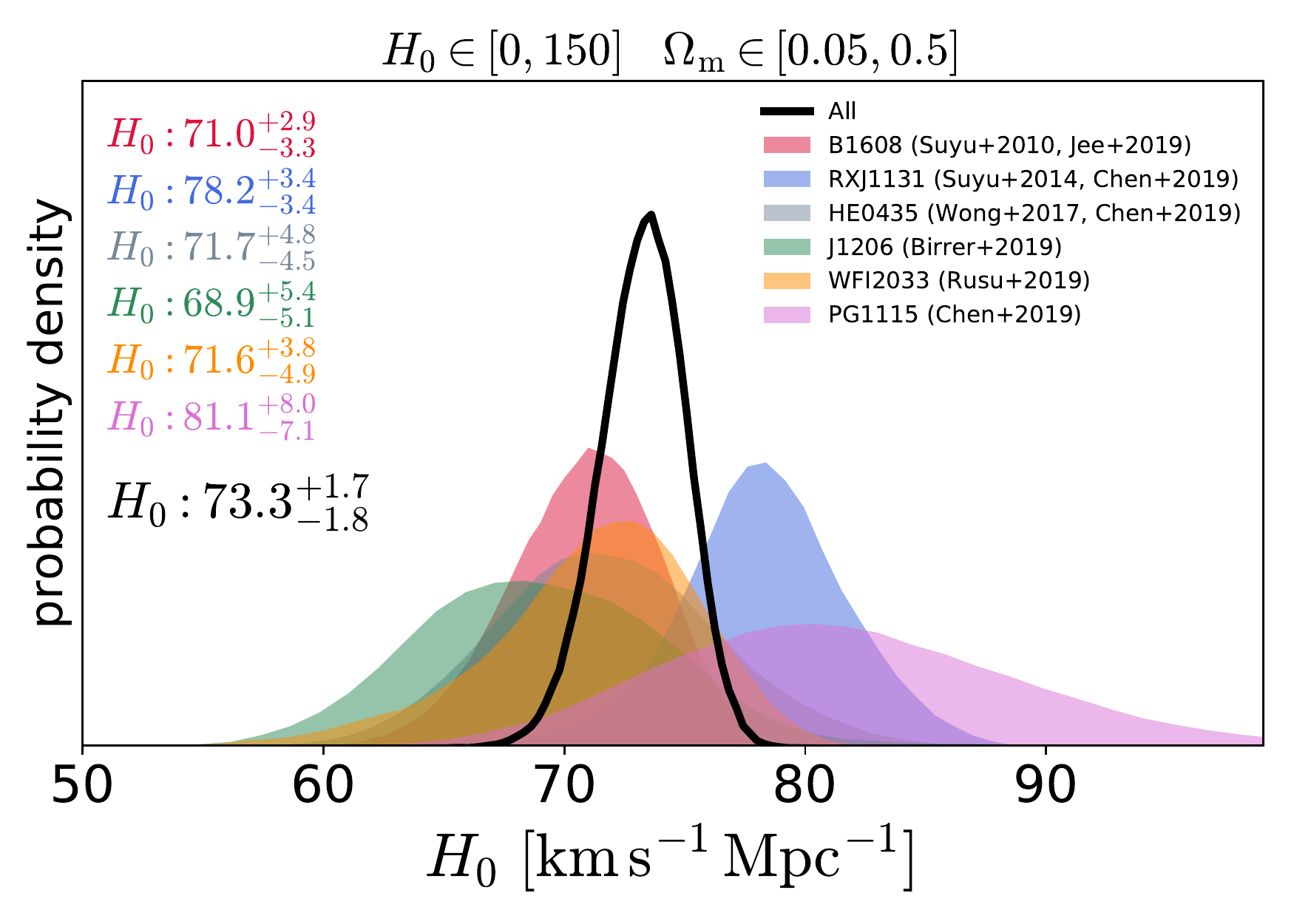}
\caption{Marginalized $H_{0}$ for a flat $\Lambda$CDM cosmology with uniform priors.  Shown are the $H_{0}$ posterior PDFs for the individual lens systems (shaded curves), as well as the combined constraint from all six systems (black line).  The median and 16th and 84th percentiles are shown in the figure legend.}
\label{fig:h0_flcdm}
\end{figure*}

Our baseline model is the flat $\Lambda$CDM cosmology with a uniform prior on $H_{0}$ in the range $[0,150]$ km s$^{-1}$ Mpc$^{-1}$ and a uniform prior on $\Om$ in the range $[0.05,0.5]$.  In Figure~\ref{fig:h0_flcdm}, we show the marginalized constraints on $H_{0}$ from each of the individual H0LiCOW lenses along with the combined constraint from all six systems.  We find $H_{0} = \ulcdm$, a $\hprec\%$ precision measurement.  We show the median and 68\% quantiles of the cosmological parameter distributions in Table~\ref{tab:cosmo_lensonly}.  This estimate is higher than the \citet{planck+2018b} CMB value ($H_{0} = 67.4 \pm 0.5~\mathrm{km~s^{-1}~Mpc^{-1}}$) by $\lenstension\sigma$ and in agreement with the latest SH0ES result ($H_{0} = 74.03 \pm 1.42~\mathrm{km~s^{-1}~Mpc^{-1}}$) from \citet{riess+2019}.

\renewcommand*\arraystretch{1.5}
\begin{table*}
\caption{Cosmological parameters for various cosmologies from time-delay cosmography only. \label{tab:cosmo_lensonly}}
\begin{minipage}{\linewidth}
\centering
\begin{tabular}{l|cccccc}
\hline
Model &
$H_{0}$ (km s$^{-1}$ Mpc$^{-1}$) &
$\Om$ &
$\OL$ or $\Ode$ &
$\Ok$ &
$w$ or $w_{0}$ &
$w_{a}$
\\
\hline
U$\Lambda$CDM &
$73.3_{-1.8}^{+1.7}$ &
$0.30_{-0.13}^{+0.13}$ &
$0.70_{-0.13}^{+0.13}$ &
$\equiv$ 0 &
$\equiv$ $-1$ &
$\equiv$ 0
\\
Uo$\Lambda$CDM &
$74.4_{-2.3}^{+2.1}$ &
$0.24_{-0.13}^{+0.16}$ &
$0.51_{-0.18}^{+0.21}$ &
$0.26_{-0.25}^{+0.17}$ &
$\equiv$ $-1$ &
$\equiv$ 0
\\
U$w$CDM &
$81.6_{-5.3}^{+4.9}$ &
$0.31_{-0.10}^{+0.11}$ &
$0.69_{-0.11}^{+0.10}$ &
$\equiv$ 0 &
$-1.90_{-0.41}^{+0.56}$ &
$\equiv$ 0
\\
U$w_{0}w_{a}$CDM &
$81.3_{-5.4}^{+5.1}$ &
$0.31_{-0.11}^{+0.11}$ &
$0.69_{-0.11}^{+0.11}$ &
$\equiv$ 0 &
$-1.86_{-0.45}^{+0.63}$ &
$-0.05_{-1.37}^{+1.45}$
\\
\hline
\end{tabular}
\\
{\footnotesize Reported values are medians, with errors corresponding to the 16th and 84th percentiles.}
\\
\end{minipage}
\end{table*}
\renewcommand*\arraystretch{1.0}

\citet{bonvin+2017} noted that the first three H0LiCOW systems showed a trend of lower lens redshift systems having a larger inferred value of $H_{0}$, but could not conclude anything due to the small sample size.  With a sample of six lenses, we see that this general trend still remains, as well as a trend of decreasing $H_{0}$ with increasing $\tdist$.  Even with six lenses, these correlations are not significant enough to conclude whether this is a real effect arising from some unknown systematic, a real physical effect related to cosmology, or just a statistical fluke (see Appendix~\ref{app:h0_trend}).  To verify that the low lens redshift systems (\rxjlens~and \pglens) can safely be combined with the other four, we compute the Bayes factor between these two groups to be $F=1.9$, indicating that there is no statistical evidence that a different set of cosmological parameters is better representing the low redshift lenses.  Nevertheless, the persistence of these trends is something to continue to examine as the sample of time-delay lenses increases in the future.

\subsection{Extensions to Flat $\Lambda$CDM, Constraints from Time-Delay Cosmography Only} \label{subsec:ext_lensonly}
Given the current tension between determinations of $H_{0}$ from CMB observations and local probes, a possibility is that the underlying cosmology that describes our Universe is more complex than the standard flat $\Lambda$CDM model.  Here, we present constraints from time-delay cosmography alone in some common single-parameter or two-parameter extensions to flat $\Lambda$CDM.  The parameter constraints for the models we test here are given in Table~\ref{tab:cosmo_lensonly}.

\subsubsection{Open $\Lambda$CDM} \label{subsubsec:olcdm_lensonly}
A simple modification to the flat $\Lambda$CDM cosmology is an open $\Lambda$CDM cosmology that allows for spatial curvature, $\Ok \ne 0$.  In this model, we have $\Om = 1-\OL-\Ok$.  We adopt uniform prior on $\Ok$ in the range $[-0.5,0.5]$, $\OL$ in the range [0,1], and require that $\Om > 0$ .   We still maintain the uniform prior on $H_{0}$ in the range $[0,150]$ km s$^{-1}$ Mpc$^{-1}$.

\begin{figure}
\includegraphics[width=0.48\textwidth]{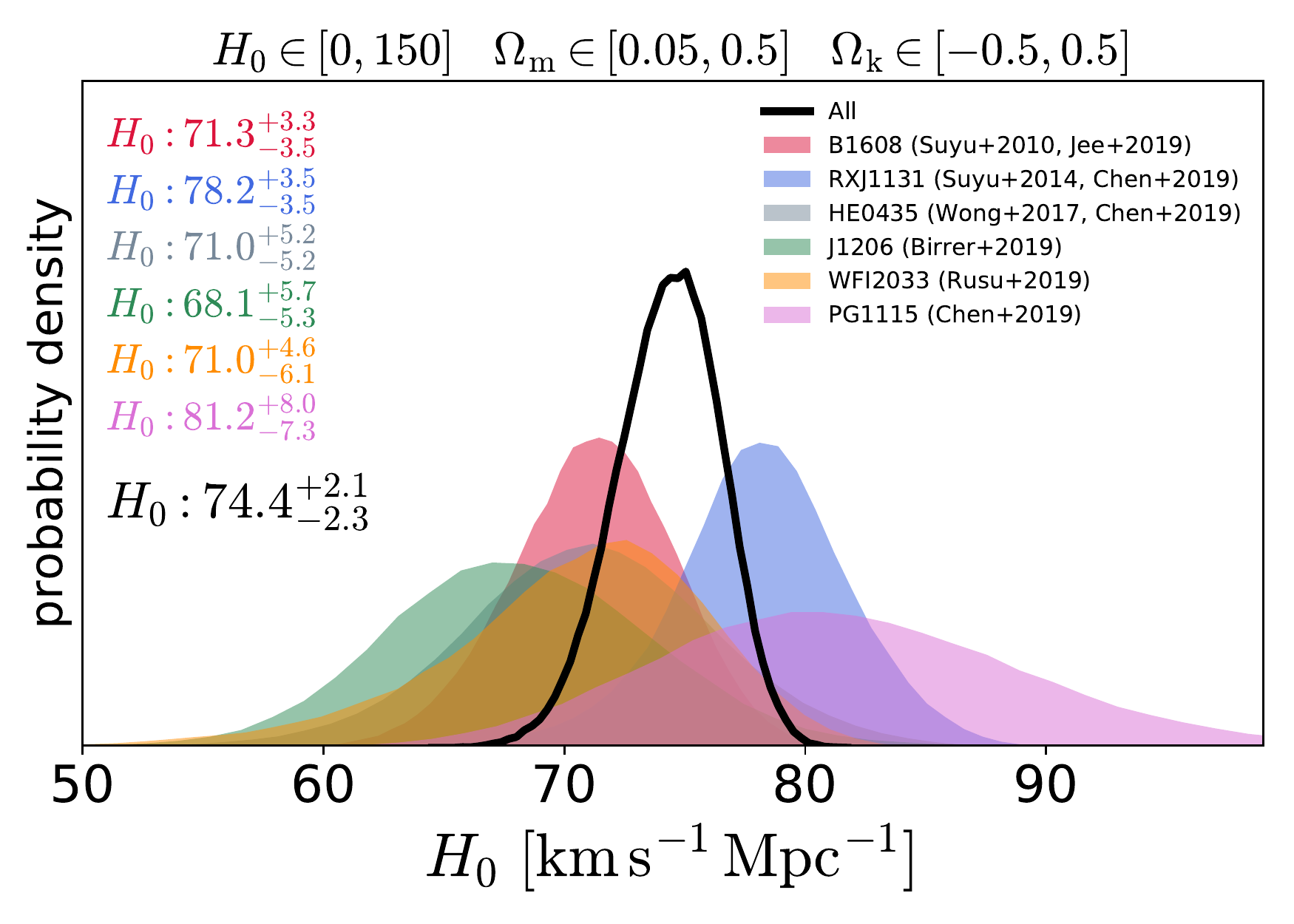}
\caption{Same as Figure~\ref{fig:h0_flcdm} for an open $\Lambda$CDM cosmology.}
\label{fig:h0_olcdm}
\end{figure}

The parameter constraints are given in Table~\ref{tab:cosmo_lensonly}.  Figure~\ref{fig:h0_olcdm} shows the marginalized constraint on $H_{0}$ in an open $\Lambda$CDM cosmology, which we find to be $H_{0} = \uolcdm$.  This is consistent with our flat $\Lambda$CDM constraint, although with a larger uncertainty.  This constraint is still inconsistent with the {\it Planck} value, indicating that allowing for spatial curvature does not resolve the tension.

\begin{figure}
\includegraphics[width=0.48\textwidth]{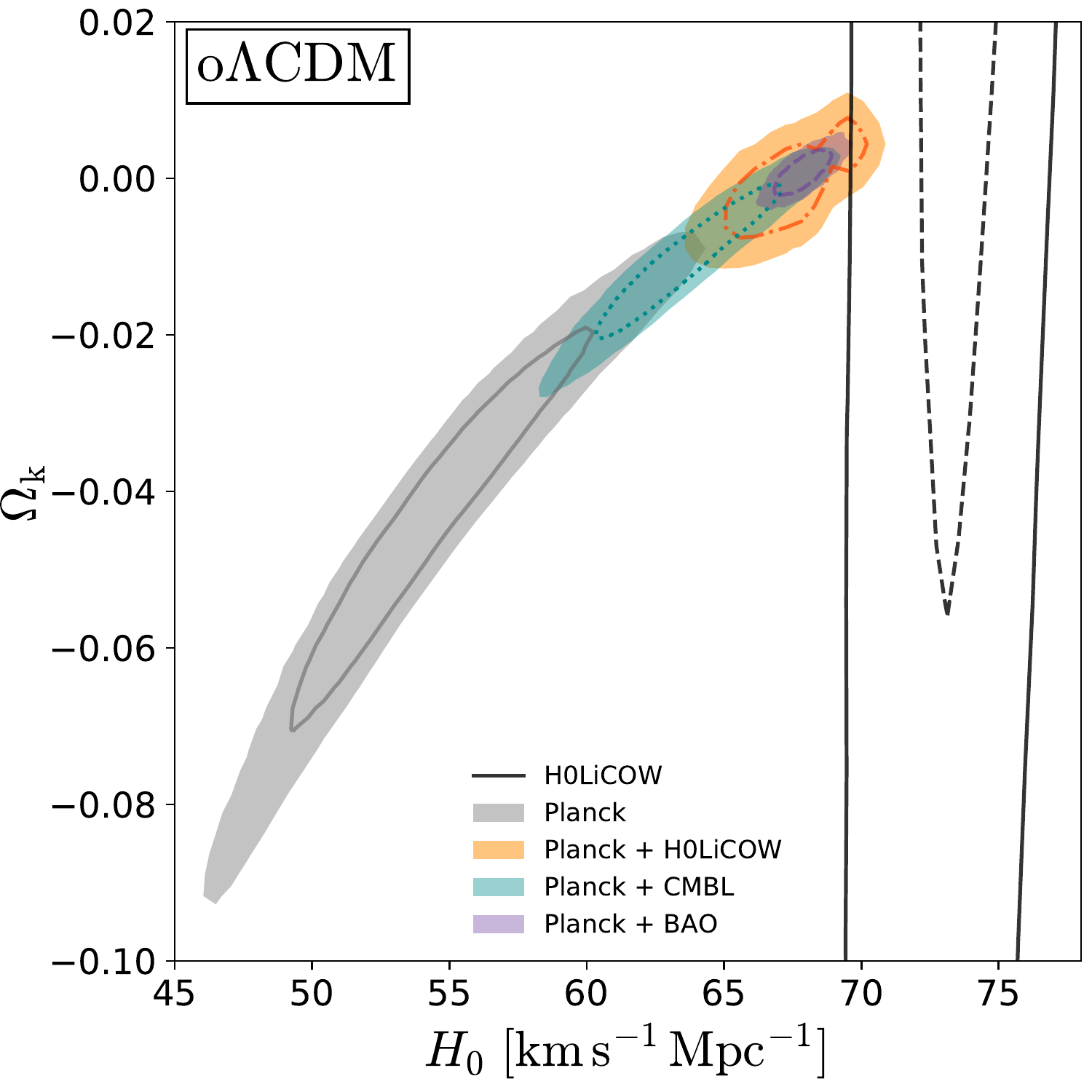}
\caption{$H_{0}$-$\Ok$ constraint for an open $\Lambda$CDM cosmology.  The black contours show the constraints from H0LiCOW alone, while the grey contours show the constraints from {\it Planck} alone.  The colored contours show constraints from {\it Planck} combined with other probes, as shown in the figure legend.  Although the H0LiCOW and {\it Planck} constraints are not consistent with each other, we show the combination here for completeness. The contour levels represent the 1$\sigma$ and 2$\sigma$ constraints.}
\label{fig:h0k_olcdm}
\end{figure}

In Figure~\ref{fig:h0k_olcdm}, we show a contour plot of the joint constraint on $H_{0}$ and $\Ok$.  The black contour is the constraint from strong lensing alone.  We see that $\Ok$ is very poorly constrained ($\Ok = \uolcdmOk$).  This is not surprising, as the time-delay distance is only weakly sensitive to $\Om$ and $\OL$, so we would expect a similar insensitivity to $\Ok$.  However, the fact that time-delay cosmography constrains $H_{0}$ very tightly indirectly imposes a tight constraint on curvature when combined with other probes.

\subsubsection{Flat $w$CDM} \label{subsubsec:wcdm_lensonly}
We consider a flat $w$CDM cosmology in which the dark energy density is not a cosmological constant, but instead is time-dependent with an equation-of-state parameter $w$.  We denote the dark energy density parameter as $\Ode = 1-\Om$.  The $w = -1$ case corresponds to flat $\Lambda$CDM with $\Ode = \OL$.  We adopt a uniform prior on $w$ in the range $[-2.5,0.5]$, keeping the same uniform priors on $H_{0}$ and $\Om$ as in the flat $\Lambda$CDM model.

\begin{figure}
\includegraphics[width=0.48\textwidth]{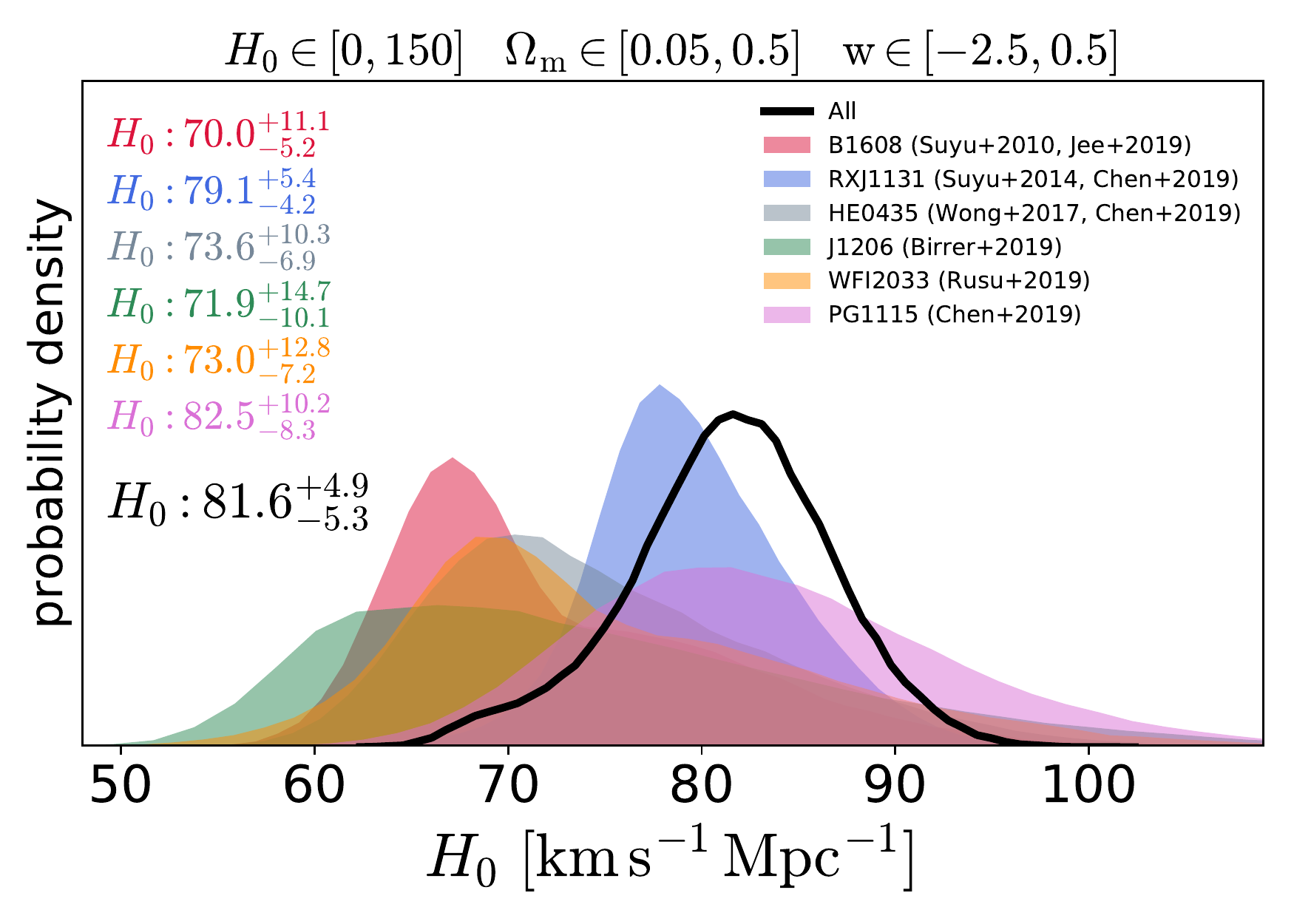}
\caption{Same as Figure~\ref{fig:h0_flcdm} for a flat $w$CDM cosmology.}
\label{fig:h0_fwcdm}
\end{figure}

We show the parameter constraints in Table~\ref{tab:cosmo_lensonly}.  In Figure~\ref{fig:h0_fwcdm}, we show the marginalized constraint on $H_{0}$ in this cosmology, which is $H_{0} = \uwcdm$.  The combined constraint on $H_{0}$ appears to be shifted to a higher value than most of the individual lenses, 
but this is due to the degeneracy between $H_0$ and $w$ and the resulting asymmetry in the PDF when projected from the higher-dimensional cosmological parameter space.  Indeed, there is a region in the two-dimensional $H_0$$-$$w$ plane in which all the individual distributions converge. That region has a higher probability density when performing the joint inference, which in turn drives the marginalized $H_0$ value higher.

\begin{figure}
\includegraphics[width=0.48\textwidth]{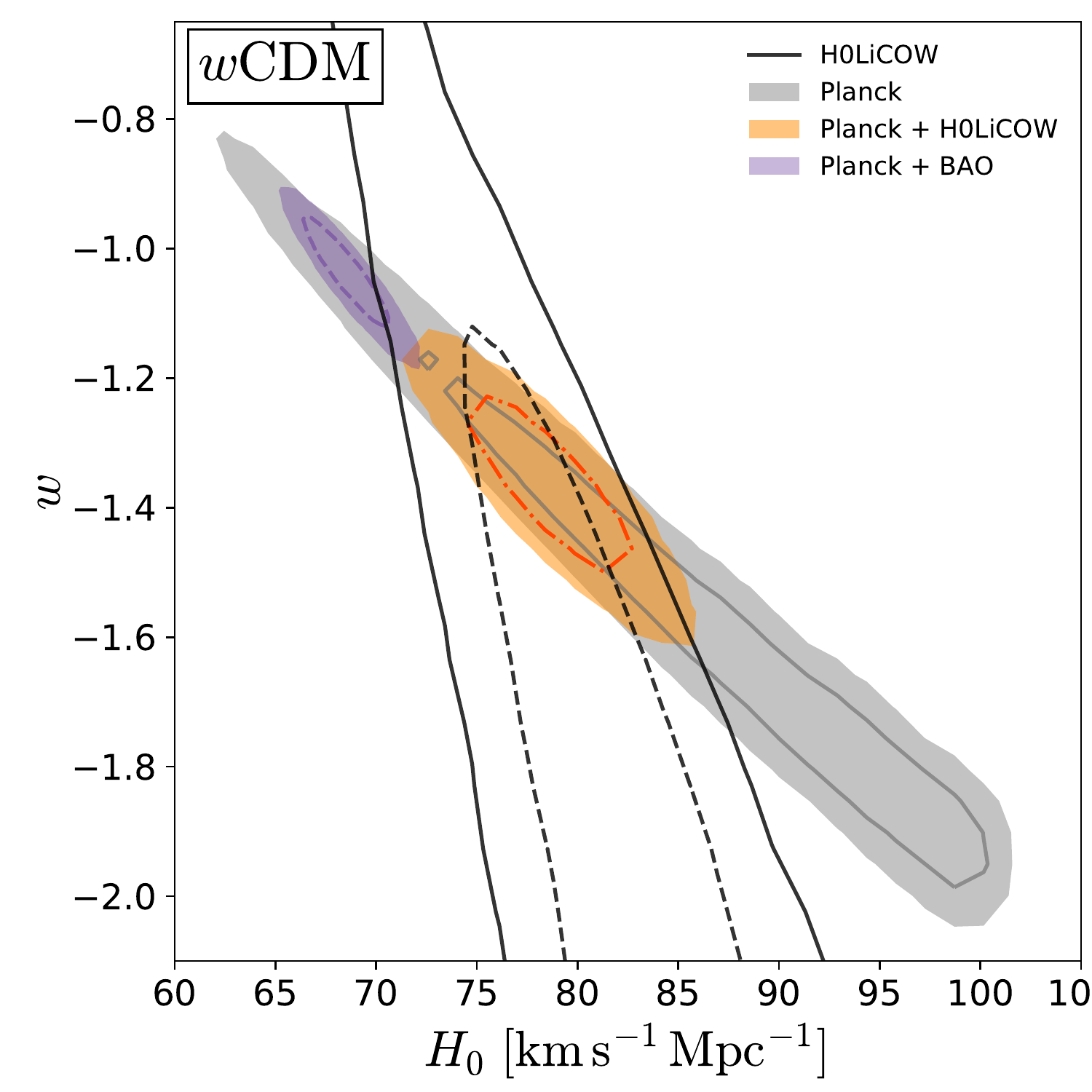}
\caption{$H_{0}$-$w$ constraint for a flat $w$CDM cosmology.  The black contours show the constraints from H0LiCOW alone, while the grey contours show the constraints from {\it Planck} alone.  The colored contours show constraints from {\it Planck} combined with other probes, as shown in the figure legend.  The contour levels represent the 1$\sigma$ and 2$\sigma$ constraints.}
\label{fig:h0w_fwcdm}
\end{figure}

In Figure~\ref{fig:h0w_fwcdm}, we show the joint distribution of $H_{0}$ and $w$.  Lensing alone does not constrain $w$ particularly well ($w = \uwcdmw$), although there is a degeneracy between $w$ and $H_{0}$, suggesting that combining our constraint with other probes may produce useful constraints.

\subsubsection{Flat $w_{0}w_{a}$CDM} \label{subsubsec:wacdm_lensonly}
The flat $w$CDM cosmology has a time-varying dark energy component with an equation of state parameter $w$.  In principle, $w$ itself could be changing with time.  We consider a flat $w_{0}w_{a}$CDM cosmology in which the dark energy equation-of-state parameter $w$ is time-varying and parameterized as $w(z) = w_{0} + w_{a} z / (1+z)$ \citep{chevallierpolarski2001,linder2003}.  We adopt a uniform prior on $w_0$ in the range $[-2.5,0.5]$ and on $w_a$ in the range $[-2,2]$, keeping the same uniform priors on $H_{0}$ and $\Om$ as in the flat $\Lambda$CDM model.

\begin{figure}
\includegraphics[width=0.48\textwidth]{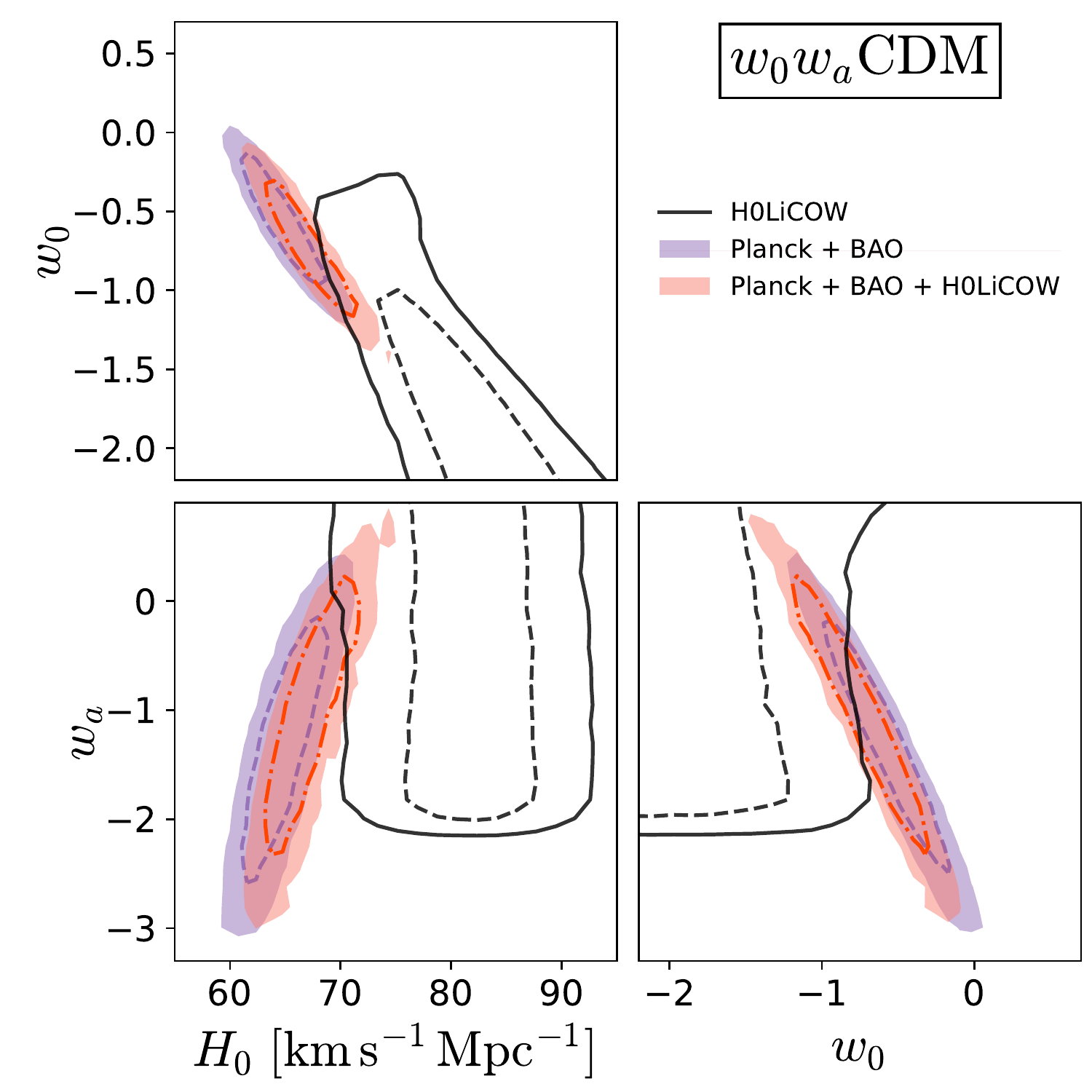}
\caption{Constraints on $H_{0}$, $w_{0}$, and $w_{a}$ for a $w_{0}w_{a}$CDM cosmology.  The colored contours show constraints from {\it Planck} combined with other probes, as shown in the figure legend (no chains for {\it Planck} alone are available for this cosmology).  The contour levels represent the 1$\sigma$ and 2$\sigma$ constraints.   H0LiCOW alone places effectively no constraints on $w_{a}$ with the resulting posterior (open black contours) spanning the prior range of $[-2,2]$.}
\label{fig:h0w0wa_w0wacdm}
\end{figure}

We show the parameter constraints in Table~\ref{tab:cosmo_lensonly}.  In Figure~\ref{fig:h0w0wa_w0wacdm}, we show the joint constraints on $H_{0}$, $w_{0}$, and $w_{a}$ from the six lenses in open black contours.  Unsurprisingly, the lenses provide little constraint on $w_0$ and effectively no constraint on $w_a$, with the posterior PDF on $w_a$ spanning the entire prior range of $[-2,2]$.  The resulting $H_0$ with a time-varying $w(z)$ is similarly high ($H_{0} = \uwwacdm$), as in the case of a flat $w$CDM cosmology, due to parameter degeneracies between $H_0$ and $w_0$.

\subsection{Extensions to Flat $\Lambda$CDM, Combinations with CMB and BAO} \label{subsec:ext_comb}
The constraints from time-delay cosmography can be combined with other probes to provide joint constraints in the open $\Lambda$CDM flat $w$CDM, and flat $w_{0}w_{a}$CDM cosmologies, as well as a number of other extensions.  Although time-delay cosmography is primarily sensitive to $H_{0}$ and only weakly dependent on other parameters, the degeneracies are such that strong lensing is highly complementary to other cosmological probes.  Here, we combine strong lensing with CMB observations from {\it Planck} \citep{planck+2018b}, sometimes including CMB weak lensing (CMBL) results, using the chains provided by the {\it Planck} team\footnote{http://pla.esac.esa.int/pla/\#cosmology}.  We can also further combine with baryon acoustic oscillation (BAO) constraints at various redshifts \citep{beutler+2011,ross+2015,alam+2017}.  All parameter constraints for the models investigated in this section are presented in Table~\ref{tab:cosmo_comb}.

\renewcommand*\arraystretch{1.5}
\setlength{\tabcolsep}{2pt}
\begin{table*}
\caption{Cosmological parameters for various cosmologies from time-delay cosmography combined with other probes.  The o$\Lambda$CDM and $w$CDM constraints can be compared with those from time-delay cosmography only (Table~\ref{tab:cosmo_lensonly}). \label{tab:cosmo_comb}}
\begin{minipage}{\linewidth}
\centering
\begin{tabular}{ll|cccccccc}
\hline
Model &
H0LiCOW + &
$H_{0}$ (km s$^{-1}$ Mpc$^{-1}$) &
$\Om$ &
$\OL$ or $\Ode$ &
$\Ok$ &
$w$ or $w_{0}$ &
$w_{a}$ &
$\mathrm{N_{eff}}$ &
$\sum \mathrm{m}_{\nu}$ (eV)
\\
\hline
o$\Lambda$CDM &
{\it Planck} &
$67.0_{-1.4}^{+1.3}$ &
$0.32_{-0.01}^{+0.01}$ &
$0.68_{-0.01}^{+0.01}$ &
$-0.002_{-0.003}^{+0.003}$ &
$\equiv$ $-1$ &
$\equiv$ 0 &
$\equiv$ 3.046 &
$\equiv$ 0.06
\\
o$\Lambda$CDM &
{\it Planck} + CMBL &
$69.2_{-1.8}^{+1.7}$ &
$0.30_{-0.01}^{+0.02}$ &
$0.70_{-0.01}^{+0.01}$ &
$0.003_{-0.004}^{+0.003}$ &
$\equiv$ $-1$ &
$\equiv$ 0 &
$\equiv$ 3.046 &
$\equiv$ 0.06
\\
o$\Lambda$CDM &
{\it Planck} + BAO &
$68.4_{-0.7}^{+0.7}$ &
$0.30_{-0.01}^{+0.01}$ &
$0.69_{-0.01}^{+0.01}$ &
$0.002_{-0.002}^{+0.002}$ &
$\equiv$ $-1$ &
$\equiv$ 0 &
$\equiv$ 3.046 &
$\equiv$ 0.06
\\
$w$CDM &
{\it Planck} &
$78.7_{-2.7}^{+2.7}$ &
$0.23_{-0.01}^{+0.02}$ &
$0.77_{-0.02}^{+0.01}$ &
$\equiv$ 0 &
$-1.36_{-0.09}^{+0.09}$ &
$\equiv$ 0 &
$\equiv$ 3.046 &
$\equiv$ 0.06
\\
$w$CDM &
{\it Planck} + BAO &
$70.3_{-1.5}^{+1.6}$ &
$0.29_{-0.01}^{+0.01}$ &
$0.71_{-0.01}^{+0.01}$ &
$\equiv$ 0 &
$-1.10_{-0.06}^{+0.06}$ &
$\equiv$ 0 &
$\equiv$ 3.046 &
$\equiv$ 0.06
\\
\hline
$\mathrm{N_{eff}}\Lambda$CDM &
{\it Planck} &
$68.8_{-1.3}^{+1.2}$ &
$0.31_{-0.01}^{+0.01}$ &
$0.69_{-0.01}^{+0.01}$ &
$\equiv$ 0 &
$\equiv$ $-1$ &
$\equiv$ 0 &
$3.20_{-0.16}^{+0.15}$ &
$\equiv$ 0.06
\\
$\mathrm{N_{eff}}\Lambda$CDM &
{\it Planck} + BAO &
$69.0_{-1.1}^{+1.1}$ &
$0.30_{-0.01}^{+0.01}$ &
$0.70_{-0.01}^{+0.01}$ &
$\equiv$ 0 &
$\equiv$ $-1$ &
$\equiv$ 0 &
$3.23_{-0.17}^{+0.17}$ &
$\equiv$ 0.06
\\
$\mathrm{m}_{\nu}\Lambda$CDM &
{\it Planck} &
$68.0_{-0.7}^{+0.7}$ &
$0.31_{-0.01}^{+0.01}$ &
$0.69_{-0.01}^{+0.01}$ &
$\equiv$ 0 &
$\equiv$ $-1$ &
$\equiv$ 0 &
$\equiv$ 3.046 &
$0.03_{-0.02}^{+0.05}$
\\
$\mathrm{m}_{\nu}\Lambda$CDM &
{\it Planck} + CMBL &
$68.0_{-0.7}^{+0.6}$ &
$0.31_{-0.01}^{+0.01}$ &
$0.69_{-0.01}^{+0.01}$ &
$\equiv$ 0 &
$\equiv$ $-1$ &
$\equiv$ 0 &
$\equiv$ 3.046 &
$0.01_{-0.02}^{+0.05}$
\\
$\mathrm{m}_{\nu}\Lambda$CDM &
{\it Planck} + BAO &
$68.0_{-0.5}^{+0.5}$ &
$0.31_{-0.01}^{+0.01}$ &
$0.69_{-0.01}^{+0.01}$ &
$\equiv$ 0 &
$\equiv$ $-1$ &
$\equiv$ 0 &
$\equiv$ 3.046 &
$0.03_{-0.02}^{+0.04}$
\\
$\mathrm{N_{eff}}\mathrm{m}_{\nu}\Lambda$CDM &
{\it Planck}  &
$69.1_{-1.3}^{+1.3}$ &
$0.30_{-0.01}^{+0.01}$ &
$0.70_{-0.01}^{+0.01}$ &
$\equiv$ 0 &
$\equiv$ $-1$ &
$\equiv$ 0 &
$3.21_{-0.17}^{+0.20}$ &
$0.03_{-0.03}^{+0.05}$
\\
$\mathrm{N_{eff}}\mathrm{m}_{\nu}\Lambda$CDM &
{\it Planck} + BAO &
$69.0_{-1.1}^{+1.0}$ &
$0.30_{-0.01}^{+0.01}$ &
$0.70_{-0.01}^{+0.01}$ &
$\equiv$ 0 &
$\equiv$ $-1$ &
$\equiv$ 0 &
$3.21_{-0.17}^{+0.16}$ &
$0.03_{-0.02}^{+0.05}$
\\
$w_{0}w_{a}$CDM &
{\it Planck} + BAO &
$67.1_{-2.5}^{+3.0}$ &
$0.32_{-0.03}^{+0.03}$ &
$0.68_{-0.03}^{+0.03}$ &
$\equiv$ 0 &
$-0.73_{-0.30}^{+0.29}$ &
$-1.02_{-0.88}^{+0.80}$ &
$\equiv$ 3.046 &
$\equiv$ 0.06
\\
\hline
\end{tabular}
\\
{\footnotesize Reported values are medians, with errors corresponding to the 16th and 84th percentiles.}
\\
\end{minipage}
\end{table*}
\setlength{\tabcolsep}{6pt}
\renewcommand*\arraystretch{1.0}

\subsubsection{Open $\Lambda$CDM} \label{subsubsec:olcdm_comb}
We test the open $\Lambda$CDM model presented in \sref{subsubsec:olcdm_lensonly} with combined constraints from time-delay cosmography and {\it Planck}.  In Figure~\ref{fig:h0k_olcdm}, the grey contours show the constraints from {\it Planck} alone, while the orange contours show the combination of {\it Planck} with time-delay cosmography.  We note that the H0LiCOW and {\it Planck} constraints are discrepant with each other.  Nonetheless, we show the joint constraints here for completeness.  The {\it Planck} data alone give a strong degeneracy between $H_{0}$ and $\Ok$, but the addition of time-delay cosmography information provides important complementary information that constrains the Universe to be nearly flat ($\Ok = -0.002_{-0.003}^{+0.003}$).  This result is also consistent with the combination of {\it Planck}+CMBL, as well as {\it Planck}+BAO.

\subsubsection{Flat $w$CDM} \label{subsubsec:wcdm_comb}
We test the flat $w$CDM model presented in \sref{subsubsec:wcdm_lensonly} with combined constraints from time-delay cosmography and {\it Planck}.  The grey contours in Figure~\ref{fig:h0w_fwcdm} show the constraints from {\it Planck} alone, while the orange contours show the combination of {\it Planck} and time-delay cosmography.  Although both {\it Planck} and H0LiCOW show degeneracies between the two parameters, the degeneracy directions are slightly different, allowing the combination of the two probes to constrain $w = 1.36_{-0.09}^{+0.09}$.  There is mild tension at the $\sim 2\sigma$ level between the {\it Planck}+H0LiCOW constraints discussed above, and {\it Planck}+BAO constraints.

\subsubsection{Flat $\Lambda$CDM with variable neutrino species and/or masses} \label{subsubsec:nnumnulcdm}
The standard model has an effective number of primordial neutrino species $\mathrm{N_{eff}} = 3.046$, but additional relativistic particles in the early Universe prior to recombination could, in principle, add to this quantity.  The sum of neutrino masses has not been precisely measured, but limits have been placed by a variety of experiments.  In our analysis thus far, we have set $\sum \mathrm{m}_{\nu} = 0.06$, the minimum mass allowed by neutrino oscillation experiments \citep{patrignani2016}.  We test cosmologies in which $\mathrm{N_{eff}}$ or $\sum \mathrm{m}_{\nu}$ are allowed to vary, as well as one in which both are allowed to vary.  Time-delay cosmography alone cannot constrain either quantity, but combining it with other probes can help to break degeneracies.

\begin{figure}
\includegraphics[width=0.48\textwidth]{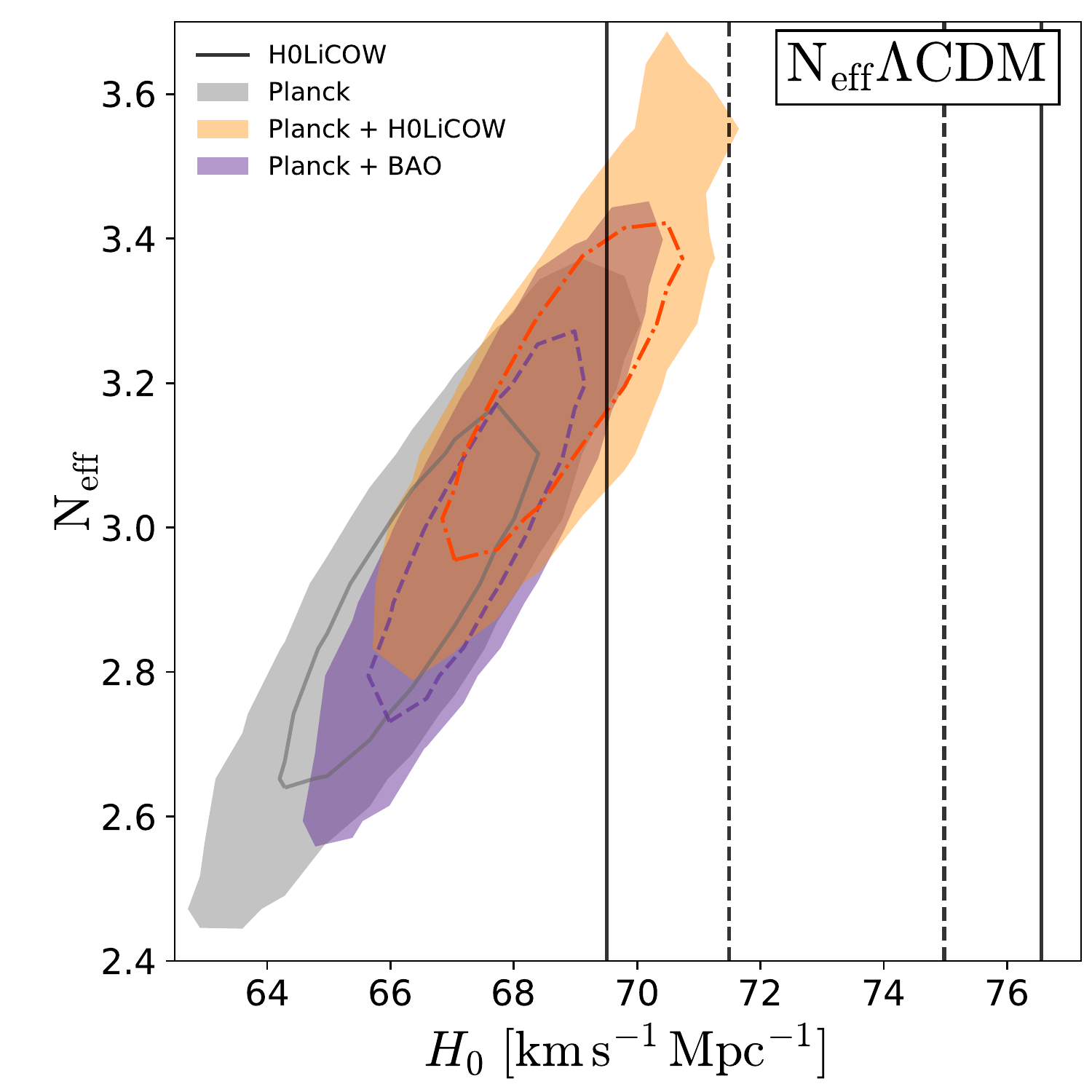}
\caption{$H_{0}$-$\mathrm{N_{eff}}$ constraint for a $\mathrm{N_{eff}}\Lambda$CDM cosmology.  The black contours show the constraints from H0LiCOW alone, while the grey contours show the constraints from {\it Planck} alone.  The colored contours show constraints from {\it Planck} combined with other probes, as shown in the figure legend.  The contour levels represent the 1$\sigma$ and 2$\sigma$ constraints.}
\label{fig:h0nnu_nnulcdm}
\end{figure}

\begin{figure}
\includegraphics[width=0.48\textwidth]{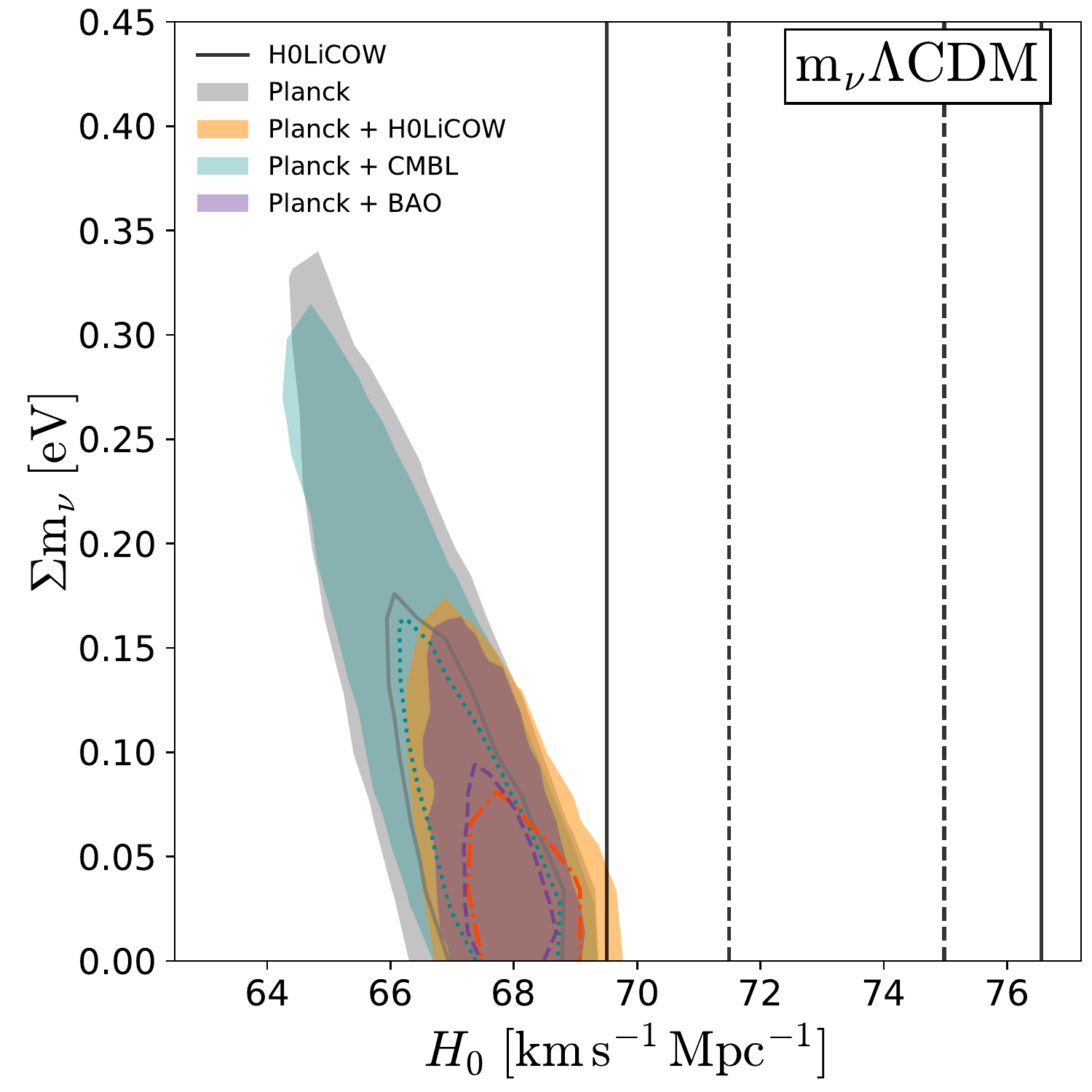}
\caption{$H_{0}$-$\sum \mathrm{m}_{\nu}$ constraint for a $\mathrm{m}_{\nu}\Lambda$CDM cosmology.  The black contours show the constraints from H0LiCOW alone, while the grey contours show the constraints from {\it Planck} alone.  The colored contours show constraints from {\it Planck} combined with other probes, as shown in the figure legend.  The contour levels represent the 1$\sigma$ and 2$\sigma$ constraints.}
\label{fig:h0mnu_mnulcdm}
\end{figure}

\begin{figure}
\includegraphics[width=0.48\textwidth]{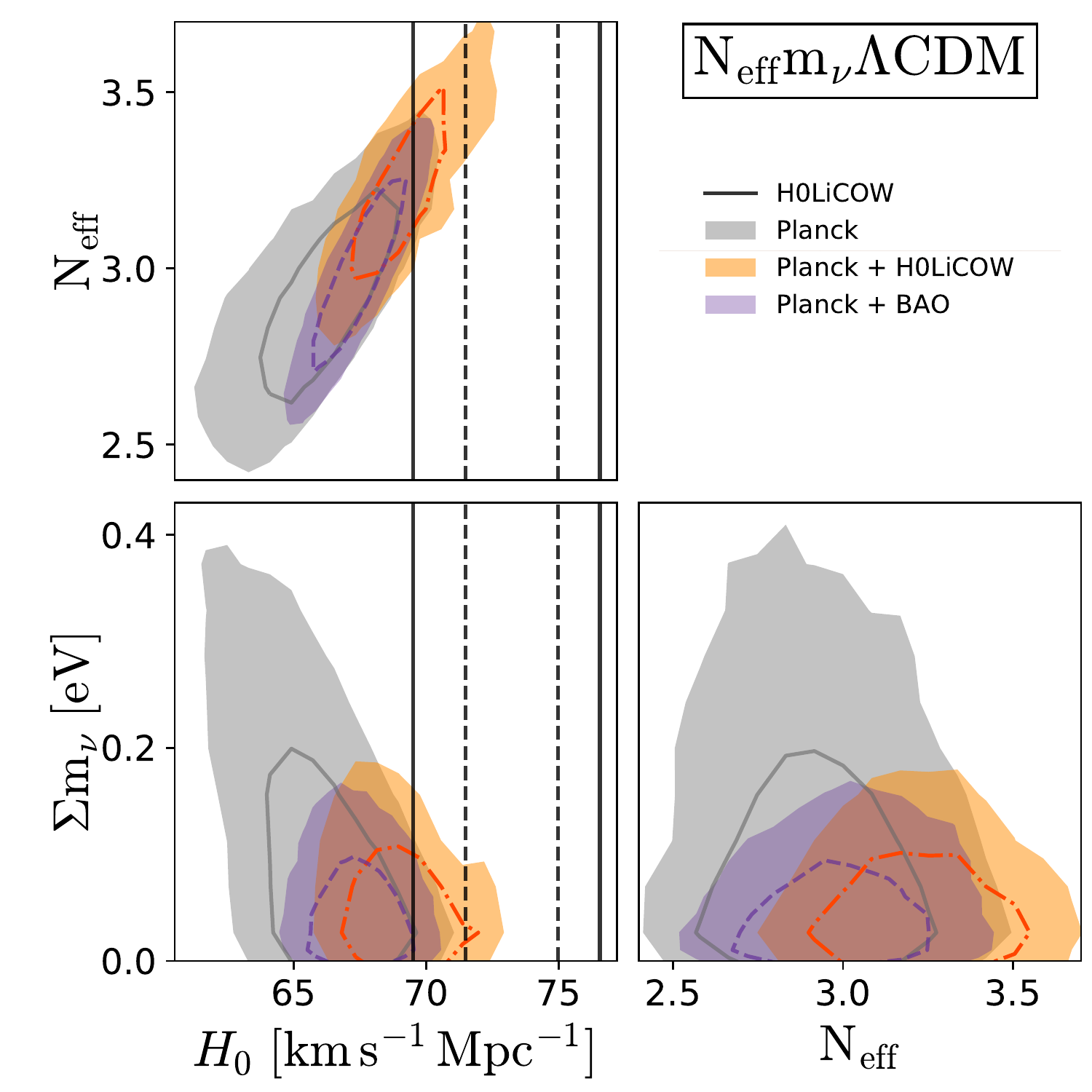}
\caption{Constraints on $H_{0}$, $\mathrm{N_{eff}}$, and $\sum \mathrm{m}_{\nu}$ for a $\mathrm{m}_{\nu}\Lambda$CDM cosmology.  The black contours show the constraints from H0LiCOW alone, while the grey contours show the constraints from {\it Planck} alone.  The colored contours show constraints from {\it Planck} combined with other probes, as shown in the figure legend.  The contour levels represent the 1$\sigma$ and 2$\sigma$ constraints.}
\label{fig:h0nnumnu_nnumnulcdm}
\end{figure}

Figure~\ref{fig:h0nnu_nnulcdm} shows a contour plot of $H_{0}$ and $\mathrm{N_{eff}}$ (when $\mathrm{N_{eff}}$ is allowed to vary).  Figure~\ref{fig:h0mnu_mnulcdm} shows a contour plot of $H_{0}$ and $\sum \mathrm{m}_{\nu}$ (when $\sum \mathrm{m}_{\nu}$ is allowed to vary).  In a model in which both quantities are allowed to vary, we show the parameter constraints in Figure~\ref{fig:h0nnumnu_nnumnulcdm}.  This model gives the combined constraints of $\mathrm{N_{eff}} = 3.21_{-0.17}^{+0.16}$ and $\sum \mathrm{m}_{\nu} = 0.03_{-0.03}^{+0.05}$, consistent with the standard model values.  Although we see that the H0LiCOW and {\it Planck} constraints in these cosmologies are somewhat in tension, we provide the joint constraints here for completeness.

\subsubsection{Flat $w_{0}w_{a}$CDM} \label{subsubsec:w0walcdm}
We consider the same flat $w_0$$w_a$CDM cosmology as in Section \ref{subsubsec:wacdm_lensonly}, now including {\it Planck} and BAO as external data sets.  The joint constraints on $H_{0}$, $w_{0}$, and $w_{a}$ are shown in Figure~\ref{fig:h0w0wa_w0wacdm}.  The combination of H0LiCOW, {\it Planck}, and BAO constraints finds $w_{0} = -0.73_{-0.30}^{+0.29}$ and $w_{a} = -1.02_{-0.88}^{+0.80}$.  Time-delay cosmography does not add much information due to the large degeneracies.

\subsection{Calibrating Type Ia Supernovae Distances with Time-Delay Cosmography} \label{subsec:snelenses}
The distance ladder method uses local distance indicators (e.g., Cepheid variables, TRGB) to calibrate the absolute distances to type Ia SNe.  In principle, any absolute distance measurement can be used to anchor the distance scale to SNe.  The inverse distance ladder method uses distances measured from baryon acoustic oscillations (BAO) in this way \citep[e.g.,][]{aubourg+2015,cuesta+2015,macaulay+2019}.

It is also possible to anchor SNe distances using either angular diameter distances ($\Dd$) to lens galaxies \citep[e.g.,][]{jee+2019,wojtakagnello2019} or $\tdist$ \citep[e.g.,][]{collett+2019,liao+2019,taubenberger+2019}.  This can be used as a complementary probe of $H_{0}$ or the sound horizon, $r_{s}$ \citep{arendse+2019a,arendse+2019b}, that is nearly insensitive to the assumed cosmology.

We follow the methodology of  \citet{taubenberger+2019}, using the combined $\tdist$ and $\Dd$ measurements from the six lenses analyzed by H0LiCOW (\sref{subsec:tddist}) to anchor measurements of type Ia SNe from the ``joint light-curve analysis" (JLA) sample of \citet{betoule+2014}.  For comparison, we also consider the Pantheon type Ia SNe sample \citep{scolnic+2018}.  We use the MontePython v3.1 MCMC sampling package \citep{audren+2013, brinckmann+2018}, its associated CLASS code \citep{lesgourgues+2011}, and the JLA and Pantheon SNe samples implemented within MontePython, to combine with our lensing distance measurements and sample cosmological parameters.  

We consider six cosmological models that are listed in part of Table \ref{tab:cosmo_models}: $\Lambda$CDM,  o$\Lambda$CDM, $w$CDM, $w_{0}w_{a}$CDM, o$w$CDM and o$w_{0}w_{a}$CDM.  For the first four cosmological models, we adopt the same uniform prior ranges for the cosmological parameters as in the top part of Table \ref{tab:cosmo_models} (i.e., same priors as U$\Lambda$CDM, Uo$\Lambda$CDM, U$w$CDM and U$w_{0}w_{a}$CDM, respectively).  For o$w$CDM and o$w_{0}w_{a}$CDM, we adopt the same priors for $w$ and $\{w_0,w_a\}$ as those in U$w$CDM and U$w_{0}w_{a}$CDM, respectively, with the remaining parameters ($H_0$, $\Om$, $\Ok$ and $\Ode$) having the same priors as those in o$\Lambda$CDM.

\begin{figure}
\includegraphics[width=0.48\textwidth]{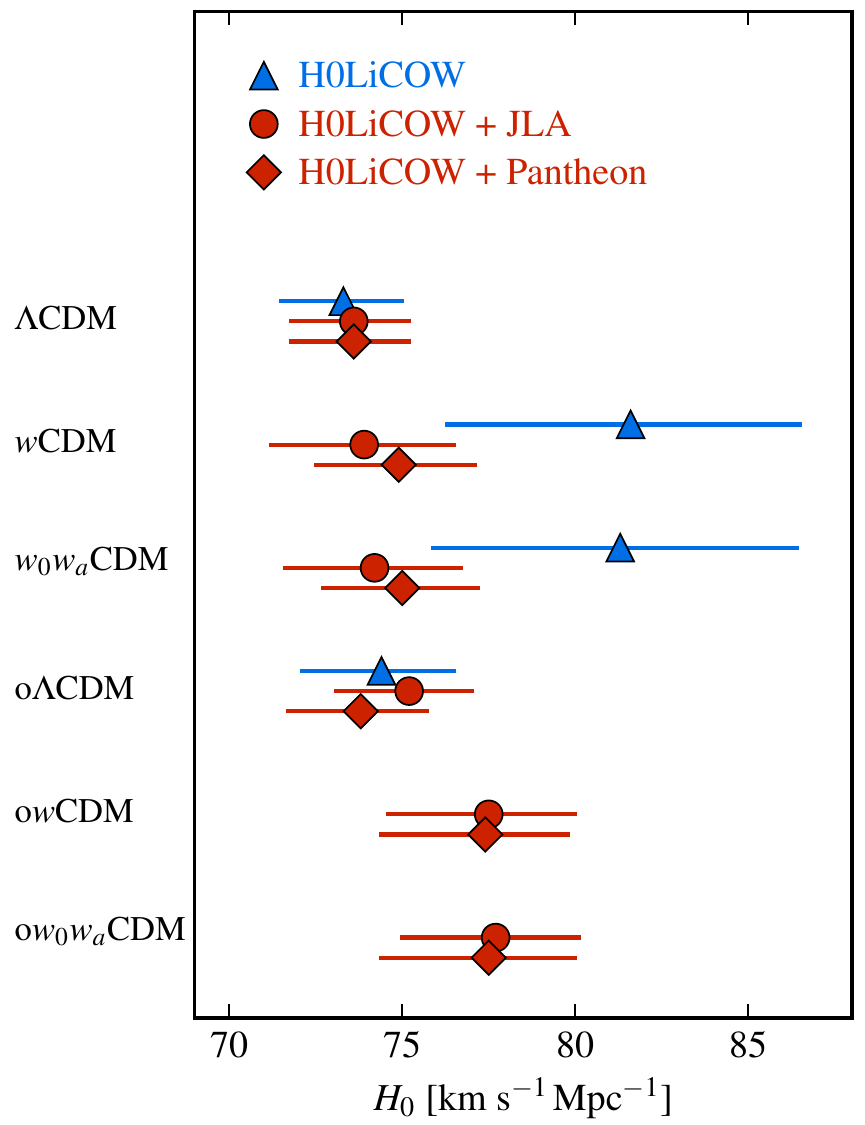}
\caption{Constraints on $H_0$ in various cosmological models (described in Table \ref{tab:cosmo_models}) from the six H0LiCOW lenses (blue triangles, Table \ref{tab:cosmo_lensonly}) and from the combination of lenses and type Ia SNe (Table \ref{tab:lens_sne1a}) using the JLA (red circles) and Pantheon (red diamonds) samples.  The points represent the median, with the error bar showing the 16th and 84th percentiles.  In the o$w$CDM and o$w_0$$w_a$CDM cosmologies, the cosmological parameter samplings of the lenses only (without SNe) do not converge due to the multiple ill-constrained parameters, and the $H_0$ values are thus not reported.  By anchoring the type Ia SNe distance scale with the lensing distances, the inferred $H_0$ is less sensitive to cosmological model assumptions, in comparison to the constraints from lenses alone.}
\label{fig:lens_sne1a_compare}
\end{figure}

Table~\ref{tab:lens_sne1a} shows the results for the six cosmologies tested, for the JLA sample (top) and the Pantheon sample (bottom).  The median values of $H_0$ from the JLA sample are within $1.5\,\kmsMpc$ of those of the Pantheon sample.  In comparison to the $H_0$ constraints in \sref{subsec:ext_lensonly}, particularly for the $w$CDM and $w_0w_a$CDM models where $H_0$ is highly degenerate with $w$, $w_0$ and $w_a$, the $H_0$ from the lenses and SNe in Table~\ref{tab:lens_sne1a} are less sensitive to cosmological models, as shown in Figure~\ref{fig:lens_sne1a_compare}.  For the six cosmological models probed in Table~\ref{tab:lens_sne1a}, the median $H_0$ values range from $\sim$73$-$78 $\mathrm{km~s^{-1}~Mpc^{-1}}$, irrespective of the SNe sample.  The tension with {\it Planck} in flat $\Lambda$CDM is still $> 3\sigma$, similar to our result from time-delay cosmography alone.

\renewcommand*\arraystretch{1.5}
\begin{table*}
\caption{Cosmological parameters for various cosmologies when anchoring type Ia supernovae distances with distances from time-delay cosmography. \label{tab:lens_sne1a}}
\begin{minipage}{\linewidth}
\centering
\begin{tabular}{l|cccccc}
\hline
\multicolumn{7}{c}{JLA sample \citep{betoule+2014} and H0LiCOW}
\\
\hline
Model &
$H_{0}$ (km s$^{-1}$ Mpc$^{-1}$) &
$\Om$ &
$\OL$ or $\Ode$ &
$\Ok$ &
$w$ or $w_{0}$ &
$w_{a}$
\\
\hline
$\Lambda$CDM &
$73.6_{-1.8}^{+1.6}$ &
$0.30_{-0.03}^{+0.03}$ &
$0.70_{-0.03}^{+0.03}$ &
$\equiv$ 0 &
$\equiv$ $-1$ &
$\equiv$ 0
\\
$w$CDM &
$73.9_{-2.7}^{+2.6}$ &
$0.31_{-0.10}^{+0.08}$ &
$0.69_{-0.08}^{+0.10}$ &
$\equiv$ 0 &
$-1.05_{-0.28}^{+0.23}$ &
$\equiv$ 0
\\
$w_{0}w_{a}$CDM &
$74.2_{-2.6}^{+2.5}$ &
$0.34_{-0.10}^{+0.08}$ &
$0.66_{-0.08}^{+0.10}$ &
$\equiv$ 0 &
$-1.04_{-0.28}^{+0.22}$ &
$-0.43_{-1.07}^{+1.10}$
\\
o$\Lambda$CDM &
$75.2_{-2.1}^{+1.8}$ &
$0.18_{-0.06}^{+0.08}$ &
$0.53_{-0.09}^{+0.12}$ &
$0.29_{-0.19}^{+0.14}$ &
$\equiv$ $-1$ &
$\equiv$ 0
\\
o$w$CDM &
$77.5_{-2.9}^{+2.5}$ &
$0.22_{-0.07}^{+0.07}$ &
$0.41_{-0.08}^{+0.14}$ &
$0.37_{-0.17}^{+0.09}$ &
$-1.46_{-0.53}^{+0.42}$ &
$\equiv$ 0
\\
o$w_{0}w_{a}$CDM &
$77.7_{-2.8}^{+2.4}$ &
$0.23_{-0.06}^{+0.07}$ &
$0.40_{-0.07}^{+0.13}$ &
$0.38_{-0.16}^{+0.09}$ &
$-1.49_{-0.52}^{+0.44}$ &
$-0.16_{-1.24}^{+1.24}$
\\
\hline
\hline
\multicolumn{7}{c}{Pantheon sample \citep{scolnic+2018} and H0LiCOW}
\\
\hline
Model &
$H_{0}$ (km s$^{-1}$ Mpc$^{-1}$) &
$\Om$ &
$\OL$ or $\Ode$ &
$\Ok$ &
$w$ or $w_{0}$ &
$w_{a}$
\\
\hline
$\Lambda$CDM &
$73.6_{-1.8}^{+1.6}$ &
$0.30_{-0.02}^{+0.02}$ &
$0.70_{-0.02}^{+0.02}$ &
$\equiv$ 0 &
$\equiv$ $-1$ &
$\equiv$ 0
\\
$w$CDM &
$74.9_{-2.4}^{+2.2}$ &
$0.35_{-0.06}^{+0.05}$ &
$0.65_{-0.05}^{+0.06}$ &
$\equiv$ 0 &
$-1.17_{-0.22}^{+0.19}$ &
$\equiv$ 0
\\
$w_{0}w_{a}$CDM &
$75.0_{-2.3}^{+2.2}$ &
$0.37_{-0.07}^{+0.05}$ &
$0.63_{-0.05}^{+0.07}$ &
$\equiv$ 0 &
$-1.15_{-0.23}^{+0.19}$ &
$-0.55_{-1.01}^{+1.22}$
\\
o$\Lambda$CDM &
$73.8_{-2.1}^{+1.9}$ &
$0.28_{-0.07}^{+0.07}$ &
$0.67_{-0.11}^{+0.11}$ &
$0.04_{-0.17}^{+0.18}$ &
$\equiv$ $-1$ &
$\equiv$ 0
\\
o$w$CDM &
$77.4_{-3.0}^{+2.5}$ &
$0.28_{-0.06}^{+0.07}$ &
$0.45_{-0.10}^{+0.17}$ &
$0.26_{-0.21}^{+0.16}$ &
$-1.52_{-0.52}^{+0.39}$ &
$\equiv$ 0
\\
o$w_{0}w_{a}$CDM &
$77.4_{-3.0}^{+2.5}$ &
$0.28_{-0.07}^{+0.08}$ &
$0.45_{-0.09}^{+0.17}$ &
$0.26_{-0.21}^{+0.15}$ &
$-1.50_{-0.47}^{+0.40}$ &
$-0.31_{-1.17}^{+1.31}$
\\
\hline
\end{tabular}
\\
{\footnotesize Reported values are medians, with errors corresponding to the 16th and 84th percentiles.}
\\
\end{minipage}
\end{table*}
\renewcommand*\arraystretch{1.0}

We also repeat the JLA analysis using only the marginalized $P(\tdist)$ from the six lenses as constraints, i.e., omitting the information from $\Dd$.  This allows us to assess the information content on $H_0$ from the lensing distances and the added value of measuring $\Dd$ in addition to $\tdist$. Using only $P(\tdist)$, we find that the $H_0$ values are very similar to values in Table~\ref{tab:lens_sne1a} (within $0.5\,\kmsMpc$), and the uncertainties are also only slightly larger (by at most $0.5\,\kmsMpc$).  Therefore, most of the cosmological information, particularly $H_0$, is encapsulated in our $\tdist$ measurements that are substantially more precise than $\Dd$ measurements.  Nonetheless, future spatially resolved kinematics of the lens galaxy could help tighten the constraints on $\Dd$, providing more leverage on cosmological parameters \citep{shajib+2018,yildirim+2019}.

\section{Tension between early-Universe and late-Universe probes of $H_{0}$} \label{sec:early_late}
As the tension between different probes of $H_{0}$ began to emerge in recent years, a natural direction to look toward in order to resolve this apparent discrepancy has been to examine potential sources of systematic error in the various methods.  In addition to exploring possible systematics in the {\it Planck} analysis and those based on type Ia SNe calibrated by the distance ladder, having multiple independent probes has proven to be a crucial step in checking these results.

\begin{figure*}
\includegraphics[width=\textwidth]{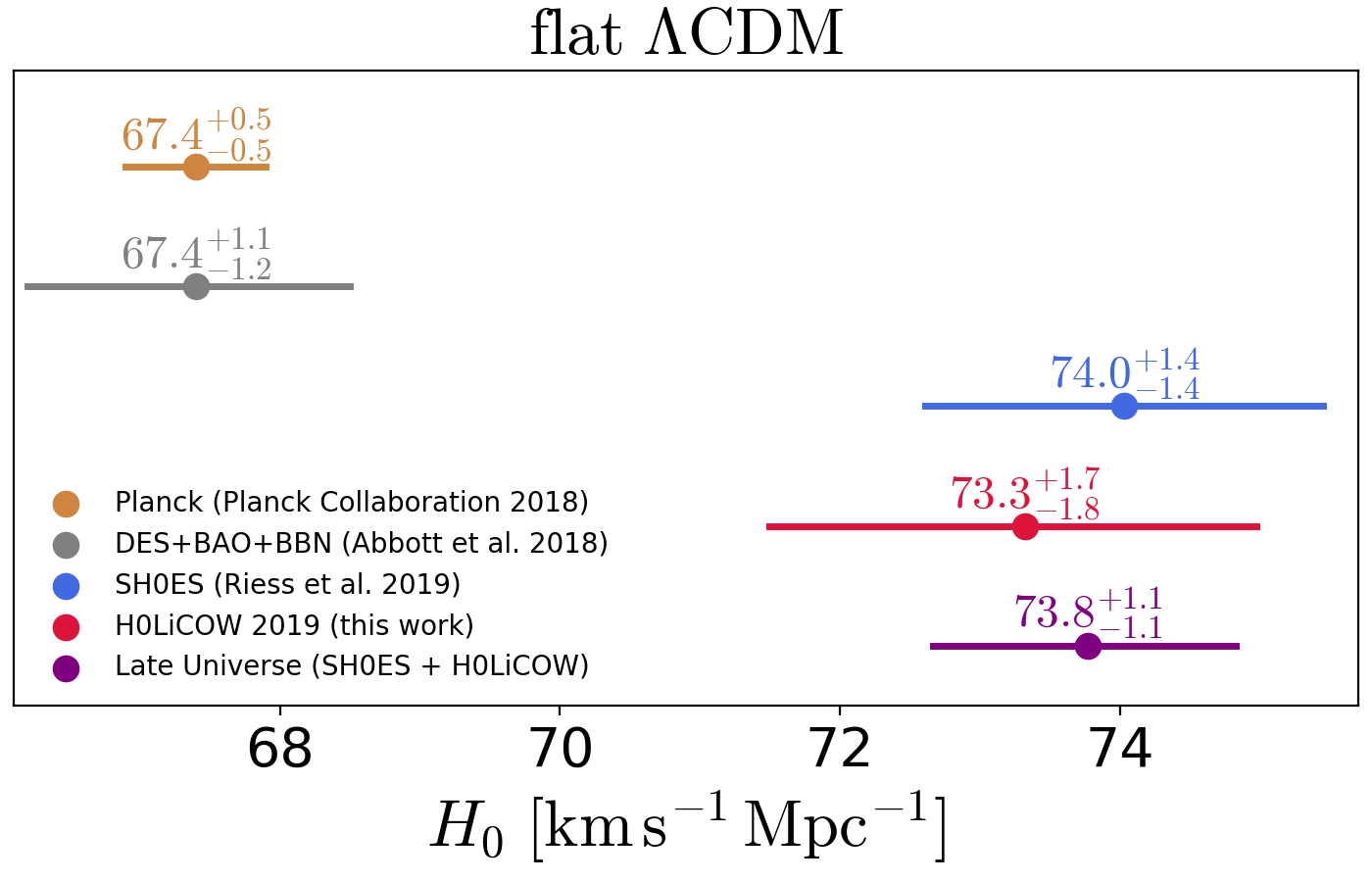}
\caption{Comparison of $H_{0}$ constraints for early-Universe and late-Universe probes in a flat $\Lambda$CDM cosmology.  The early-Universe probes shown here are from {\it Planck} \citep[orange;][]{planck+2018b} and a combination of clustering and weak lensing data, BAO, and big bang nucleosynthesis \citep[grey;][]{abbott+2018b}.  The late-Universe probes shown are the latest results from SH0ES \citep[blue;][]{riess+2019} and H0LiCOW (red; this work).  When combining the late-Universe probes (purple), we find a $\tension\sigma$ tension with {\it Planck}.}
\label{fig:h0_early_late}
\end{figure*}

The latest H0LiCOW results presented here, analyzed blindly with respect to cosmological parameters, are the most precise constraints on $H_{0}$ from time-delay cosmography to date, and are independent of both CMB probes (i.e., {\it Planck}) and other late-Universe probes such as SH0ES.  Our results for a flat $\Lambda$CDM cosmology are in $\lenstension\sigma$ tension with {\it Planck}.  In combination with the latest SH0ES result \citep{riess+2019}, we find a $\tension\sigma$ tension between late-Universe determinations of $H_{0}$ and {\it Planck} (Figure~\ref{fig:h0_early_late}).  Other independent methods anchored in the early Universe, such as the analysis of \citet{abbott+2018b} using a combination of clustering and weak lensing, BAO, and big bang nucleosynthesis (BBN), give similar results to {\it Planck}.  Although the \citet{abbott+2018b} method, as well as the inverse distance ladder, combine measurements from both the early and late Universe, the inferred $H_{0}$ is ultimately set by the sound horizon at recombination, which comes from early-Universe physics.

Given the tension shown here, it is becoming difficult to reconcile the $H_{0}$ discrepancy by appealing to systematic effects.  While systematics, especially ``unknown unknowns", still cannot be entirely ruled out and should continue to be explored, recent work has only heightened the tension.  There also appears to be a growing dichotomy when the different $H_{0}$ probes are split into those anchored by the early-Universe (i.e., CMB), which favor a lower $H_{0}$, and those based on late-Universe probes, which favor a higher $H_{0}$ \citep[e.g.,][]{verde+2019}.

As this tension between early-Universe and late-Universe probes continues to grow, we must examine potential alternatives to the standard flat $\Lambda$CDM model.  This would be a major paradigm shift in modern cosmology, requiring new physics to consistently explain all of the observational data.  We have explored some possible extensions to flat $\Lambda$CDM in Section~\ref{sec:cosmo}, including spatial curvature, time-varying dark energy \citep[e.g.,][]{divalentino+2018}, and modified neutrino physics such as sterile neutrinos \citep[e.g.,][]{wyman+2014,gelmini+2019} or self-interacting neutrinos at early times \citep[e.g.,][]{kreisch+2019}.  Other possible new physics to resolve the discrepancy include an early dark energy component to the Universe that later decays \citep[e.g.,][]{agrawal+2019,alexandermcdonough2019,aylor+2019,lin+2019,poulin+2019}, primordial non-Gaussianity \citep[e.g.,][]{adhikarihuterer2019}, decaying dark matter \citep[e.g.,][]{pandey+2019,vattis+2019}, and fifth forces \citep[e.g.,][]{desmond+2019}.

\section{Summary} \label{sec:summary}
We have combined time-delay distances and angular diameter distances from six lensed quasars in the H0LiCOW sample to achieve the highest-precision probe of $H_{0}$ to date from strong lensing time delays.  Five of the six lenses are analyzed blindly with respect to the cosmological parameters of interest.  Our main results are as follows:
\begin{itemize}
\item We find $H_{0} = \ulcdm$ for a flat $\Lambda$CDM cosmology, which is a measurement to a precision of $\hprec\%$.  This result is in agreement with the latest results from measurements of type Ia SNe calibrated by the distance ladder \citep{riess+2019} and in $\lenstension\sigma$ tension with {\it Planck} CMB measurements \citep{planck+2018b}.
\item Our constraint on $H_{0}$ in flat $\Lambda$CDM is completely independent of and complementary to the latest results from the SH0ES collaboration, so these two measurements can be combined into a late-Universe constraint on $H_{0}$.  Together, these are in tension with the best early-Universe (i.e., CMB) determination of $H_{0}$ from {\it Planck} at a significance of $\tension\sigma$.
\item We check that the lenses in our sample are statistically consistent with one another by computing Bayes factors between their $H_{0}$ PDFs.  We find that all six lenses are pairwise consistent (i.e., $F > 1$), indicating that we are not underestimating our uncertainties and are able to control systematic effects in our analysis.
\item We compute parameter constraints for cosmologies beyond flat $\Lambda$CDM.  In an open $\Lambda$CDM cosmology, we find $\Ok = \uolcdmOk$ and $H_{0} = \uolcdm$, which is still in tension with {\it Planck}, suggesting that allowing for spatial curvature cannot resolve the discrepancy.  In a flat $w$CDM cosmology, we find $H_{0} = \uwcdm$ and $w = \uwcdmw$.  In a flat $w_{0}w_{a}$CDM cosmology, we find $H_{0} = \uwwacdm$, but are unable to place meaningful constraints on $w_{0}$ and $w_{a}$.
\item We combine our constraints with {\it Planck}, including CMB weak lensing and BAO constraints.  Although time-delay cosmography is primarily sensitive to $H_{0}$, with only a weak dependence on other cosmological parameters, the constraints are highly complementary to other probes such as {\it Planck}, CMB weak lensing, and BAO.  We test the open $\Lambda$CDM and $w$CDM cosmologies, as well as cosmologies with variable effective neutrino species and/or sum of neutrino masses, and a $w$CDM cosmology with a time-varying $w$. The full parameter constraints for these models when combining H0LiCOW and {\it Planck} are given in Table~\ref{tab:cosmo_comb}.
\item We use the distance measurements from time-delay cosmography to calibrate the distance scale of type Ia SNe from the JLA and Pantheon samples.  This provides a probe of $H_{0}$ that is less dependent on the assumed cosmological model, in comparison to the constraints from lenses alone.  We find median $H_{0}$ values ranging from $\sim$73$-$78~$\mathrm{km~s^{-1}~Mpc^{-1}}$ for a range of cosmologies.  The tension with {\it Planck} for a flat $\Lambda$CDM cosmology is still $> 3\sigma$, similar to the result from time-delay cosmography alone.
\end{itemize}

Despite efforts to explore and reduce systematic errors in the various methods, the growing tension between early and late-Universe probes of $H_{0}$ has only continued to heighten.  If unresolved, this tension may force the rejection of the flat $\Lambda$CDM model in favor of new physics, which would dramatically alter our understanding of the Universe.

While considering the possibility of new physics, we are also continuing to improve the constraints from time-delay cosmography.  The current sample of six H0LiCOW systems is already the best-studied sample to date, and a number of additional lenses are being observed with high-resolution imaging \citep[e.g.,][]{shajib+2019} and monitored by COSMOGRAIL.  Moving into the future, many new lensed quasars are being discovered in large imaging surveys \citep[e.g.,][]{agnello+2015,agnello+2018a,agnello+2018b,anguita+2018,lemon+2018,lemon+2019,treu+2018}.  A sample of $\sim40$ lenses is needed to constrain $H_{0}$ to the $\sim1\%$ level \citep{jee+2016,shajib+2019}, which will be attainable in the near future.

\section*{Acknowledgements}
We thank the referee, whose suggestions were helpful in improving the clarity of this paper.
We thank Chiara Spiniello, Malte Tewes, and Ak{\i}n Y{\i}ld{\i}r{\i}m for their contributions to the H0LiCOW project.
We thank Lodovico Coccato and Johan Richard for their help with the MUSE data used as a part of this project.
We thank Aleksi Halkola for support with the {\sc Glee} lens modeling code.
We thank all the observers at the Euler, SMARTS, Mercator
and Maidanak telescopes who participated in the queue-mode
observations. H0LiCOW and COSMOGRAIL are made possible thanks to the continuous work of all observers and technical staff obtaining the monitoring observations, in particular at the Swiss Euler telescope at La Silla Observatory. The Euler telescope is supported by the Swiss National Science Foundation.
This work was supported by World Premier International Research Center Initiative (WPI Initiative), MEXT, Japan.
K.C.W. is supported in part by an EACOA Fellowship awarded by the East Asia Core Observatories Association, which consists of the Academia Sinica Institute of Astronomy and Astrophysics, the National Astronomical Observatory of Japan, the National Astronomical Observatories of the Chinese Academy of Sciences, and the Korea Astronomy and Space Science Institute.
S.H.S. thanks the Max Planck Society for support through the Max Planck Research Group.
S.H.S.~and S.T.~are supported by the European Research Council (ERC) under the European Union's Horizon 2020 research and innovation programme (grant agreement No 771776).
G.C.-F.C. acknowledges support from the Ministry of Education in Taiwan via Government Scholarship to Study Abroad (GSSA).
C.D.F. and G.C.-F.C. acknowledge support for this work from the National Science Foundation under Grant No. AST-1715611.
M.M, D.S., V.B., F.C., and O.T. are supported by the Swiss National Science Foundation (SNSF) and by the European Research Council (ERC) under the European Union's Horizon 2020 research and innovation programme (COSMICLENS: grant agreement No 787886). 
T.T. acknowledges support by the Packard Foundation through a Packard Research fellowship and by the National Science Foundation through NSF grant AST-1714953.
A.A. was supported by a grant from VILLUM FONDEN (project number 16599). This project is partially funded by the Danish council for independent research under the project ``Fundamentals of Dark Matter Structures'', DFF--6108-00470.
S.H. acknowledges support by the DFG cluster of excellence \lq{}Origin and Structure of the Universe\rq{} (\href{http://www.universe-cluster.de}{\texttt{www.universe-cluster.de}}).
A.J.S. acknowledges support by NASA through Space Telescope Science Institute grant HST-GO-15320.
L.V.E.K. is partly supported through an NWO-VICI grant (project number 639.043.308).
Based on observations made with the NASA/ESA Hubble Space Telescope, obtained at the Space Telescope Science Institute, which is operated by the Association of Universities for Research in Astronomy, Inc., under NASA contract NAS 5-26555. These observations are associated with programs HST-GO-9375, HST-GO-9744, HST-GO-10158, HST-GO-12889, and HST-14254. Support for programs HST-GO-10158 HST-GO-12889 HST-14254 was provided to members of our team by NASA through a grant from the Space Telescope Science Institute, which is operated by the Association of Universities for Research in Astronomy, Inc., under NASA contract NAS 5-26555.
This research made use of Astropy,\footnote{http://www.astropy.org} a community-developed core Python package for Astronomy \citep{astropy+2013,astropy+2018}.
This research made use of Matplotlib, a 2D graphics package used for Python \citep{hunter2007}.
This research made use of {\tt emcee}, a Python implementation of an affine invariant MCMC ensemble sampler \citep{foremanmackey+2013}.


\bibliography{cosmo6lenses}
\bibliographystyle{mnras}


\appendix

\section{Quantifying the trend of $H_{0}$ with $\zd$ and $\tdist$} \label{app:h0_trend}
\begin{figure*}
\centering
\includegraphics[width=\textwidth]{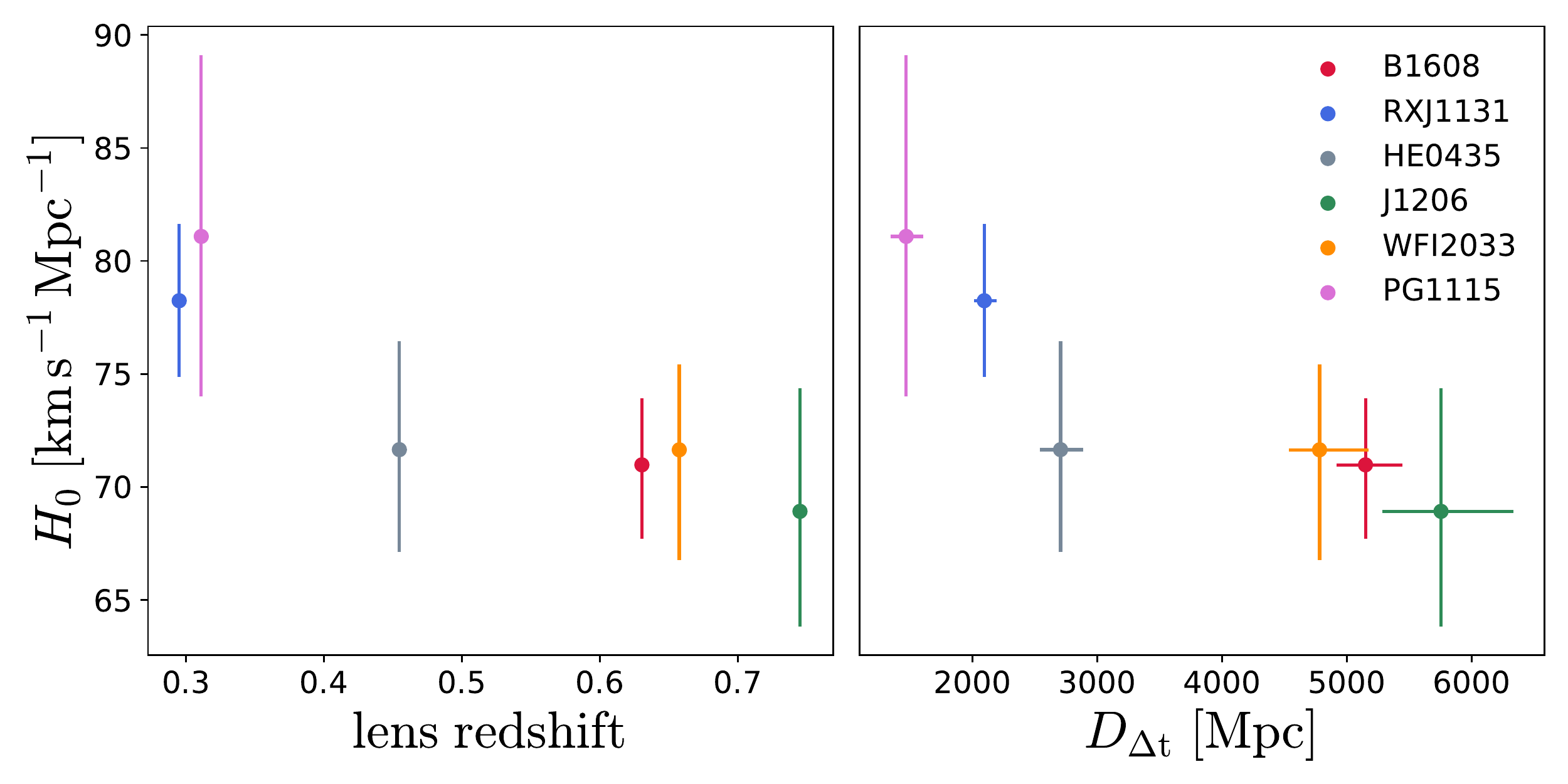}
\caption{$H_{0}$ constraints for the individual H0LiCOW lenses as a function of lens redshift (left) and time-delay distance (right).  The trend of smaller $H_{0}$ value with increasing lens redshift and with increasing $\tdist$ has significance levels of 1.9$\sigma$ and 1.8$\sigma$, respectively.
}
\label{fig:h0_trend}
\end{figure*}

There is an apparent trend of decreasing $H_{0}$ inferred from the individual lenses as a function of increasing lens redshift, which is shown in the left panel of Figure~\ref{fig:h0_trend}.  The significance of this correlation can be assessed by testing against the null hypothesis, in which the measured $H_{0}$ values are uncorrelated with the lens redshift. To do so, we draw sets of six mock $H_{0}$ values, using each lens' own uncertainty probability distribution centered around the median joint inference obtained in flat $\Lambda$CDM ($H_{0} = \ulcdm$). We then fit a linear regression through each mock set. Associating a weight to the mock value from each lens is done according to the following scheme: we first rescale the uncertainties' probability distributions so that their maximum value equal one. Next, we compute the area under each rescaled distribution, then rescale the areas by their median. Finally, we take the inverse square of the rescaled areas as weight for each mock measurement. The slope of the regression is taken as our measurement of the correlation. We create $10^5$ sets of mock values, for which the distribution of the measured slopes is centered around zero, as expected for a null hypothesis. We fit the same kind of linear regression through the data, for which we find a negative slope.
We find that the slope of the data falls 1.9$\sigma$ away from the mock slope distribution.  We also observe a correlation between $\tdist$ and $H_0$ (Figure~\ref{fig:h0_trend}, right panel) that deviates from the null hypothesis at a similar significance level of $1.8\sigma$.

\end{document}